\documentclass[aip,jcp,preprint]{revtex4-1}
\usepackage{amsmath}
\usepackage{braket}
\usepackage{amsfonts}
\usepackage{amssymb}
\usepackage{bm}
\usepackage{graphics}
\usepackage{natbib}
\usepackage{subcaption}
\usepackage{newfloat}
\usepackage{algcompatible}
\usepackage[dvipsnames]{xcolor}

\definecolor{resonance-colour}{gray}{0.95} 

\usepackage{tikz}
\usetikzlibrary{arrows}
\usetikzlibrary{decorations.pathmorphing}
\usetikzlibrary{decorations.markings}
\usetikzlibrary{decorations.pathreplacing}

\newcommand{\ddt}[1]{\frac{\mathrm{d} #1}{\mathrm{d} t}} 
\newcommand{\field}{\varepsilon} 
\newcommand{\Tr}{\text{Tr}}     
\newcommand{\super}[1]{\mathcal{#1}} 
\newcommand{\commute}[2]{[#1, #2]} 

\begin{document}

\title{Ultrashort pulse two-photon coherent control of a macroscopic phenomena: light-induced current from channelrhodopsin-2 in live brain cells}
\date{\today}
\author{Cyrille Lavigne}
\author{Paul Brumer}
\affiliation{
  Chemical Physics Theory Group, Department of Chemistry,
  and Center for Quantum Information and Quantum Control,
  University of Toronto, Toronto, Ontario, M5S 3H6, Canada}

\begin{abstract}
  Coherent control is extended to macroscopic processes under
  continuous pulsed laser irradiation. Here, this approach is used to
  analyze the experimentally measured two-photon phase control of currents
  emanating from a living brain cells expressing channelrhodopsin-2, a
  light-gated ion channel.  In particular, a mechanism is proposed that
  encompasses more than 15 orders magnitude in time, from the
  ultrafast dynamics of retinal in channelrhodopsin-2 to the slow
  dynamics of the neuron current.  Implications for other
  photochemical processes are discussed.
\end{abstract}

\maketitle

\section{Introduction}
A quantum system interacting with radiation is sensitive to both the
intensity of the exciting radiation as well as its
phase.\cite{shapiro_quantum_2012} The use of the phase content of
light to steer quantum behavior, an essential component of the
coherent control methodology, has seen numerous successful
applications.\cite{zhu_coherent_1995,hache_observation_1997,
  stievater_rabi_2001,brinks_visualizing_2010} However, past
experimental and theoretical investigations have primarily focused on
microscopic phenomena, that is on the behavior of molecules after
interaction with few pulses of light. Yet the repeated application
shaped pulses of light to a macroscopic system can indeed lead to
macroscopic phase control, as shown in a recent control experiment
performed on brain tissue.\cite{paul_coherent_2017}

Specifically, Boppart's group demonstrated that the current emanating
from living brain cells expressing channelrhodopsin-2 (ChR2), a
light-gated ion channel,\cite{lorenz-fonfria_channelrhodopsin_2014,
schneider_biophysics_2015} is sensitive to the phase of a two-photon
excitation.  In this experiment neurons are exposed to a train of
ultrashort chirped pulses over a macroscopically long time (1200 ms).
The near-infrared pulses impinging on the neurons photoisomerize
retinal in ChR2, thus activating it.  Target neurons then evoke an
electrical current.  Critically, the measured current is found to be
sensitive to both the sign and magnitude of the pulse chirp, a pure
phase control effect.\cite{arango_communication_2013}

In contrast with previous work on the control of biological
molecules,\cite{herek_quantum_2002,prokhorenko_coherent_2006} the
demonstrated phase control is macroscopic in both magnitude and time.
The measured currents are evoked by the cell over a full second of
exposure to light.  The dynamics of the electrical response of neurons
occur over the millisecond
timescale,\cite{lorenz-fonfria_channelrhodopsin_2014} the
excitation and isomerization dynamics of retinal in ChR2 are femto to
picosecond
processes,\cite{schneider_biophysics_2015,johnson_local_2015} and the
separation between individual laser pulses is on the scale of
nanoseconds.  Hence, phase control and its effects were found to span a
time range of over 15 orders of magnitude.

This experiment provides a demonstration of how weak multiphoton
control from ultrafast pulses can lead to macroscopically measurable
effects.  Similar control could be applied to other isomerization
processes\cite{izquierdo-serra_two-photon_2014,
  carroll_two-photon_2015,wei_two-photon_2019}
as well as to photochemical reactions in the condensed
phase.\cite{scaiano_photochemistry_1988,
  pastirk_quantum_1998,potapov_two-photon_2001}
Quantum control, and in particular phase control, has important
potential applications to optogenetics and photochemistry.  For example,
phase control provides a means of selectively exciting neighbouring
molecules that overlap
spectrally.\cite{pastirk_selective_2003,
  schelhas_advantages_2006,ogilvie_use_2006,
  tkaczyk_control_2009,isobe_multifarious_2009,
  brenner_two-photon_2013}
In scattering media, the exciting field will acquire a spatially
dependent phase; phase control is thus also a potential route to
superresolution targeting of
excitations.\cite{katz_focusing_2011,mounaix_spatiotemporal_2016}

The methodology of coherent control provides a rigorous and
physically-motivated description of phase coherence and phase
interference in light-matter interactions.\cite{shapiro_quantum_2012}
Physical conditions on interference can be used to devise new control
schemes, i.e. methods for phase
controllability.\cite{shapiro_quantum_1995,spanner_communication_2010}
From a phenomenological point of view, coherent control schemes are
also potential mechanisms for experimental phase control obtained in
any number of ways, e.g. chirp control\cite{hoki_mechanisms_2005} or
adaptive feedback.\cite{lavigne_interfering_2017}

In this article we examine the experiments in
Ref.~\onlinecite{paul_coherent_2017} in order to expose the underlying
coherent control mechanism.  To do so we first develop a theoretical
description of two-photon coherent control with ultrashort pulses.\cite{*[{The case of two-photon coherent control using  continuous-wave lasers is described in e.g., }] [{}]  chen_coherent_1992,*chen_theory_1993,*chen_multiproduct_1993}  
Three control schemes are then proposed as feasible mechanisms for the
phase-controlled current generation in live brain cells, based on an
analysis of possible interference pathways.  Spectroscopic means are
then described to distinguish and characterize each of the control
mechanisms. Importantly, we show that microscopic interference effects
due to repeated coherent interactions can accumulate to produce a
macroscopic coherent control result, even in the presence of fast
decoherence and dissipation.  We demonstrate this in a computational
study, where a quantum mechanical (minimal) model of
retinal\cite{hahn_quantum-mechanical_2000,
  hahn_ultrafast_2002,balzer_mechanism_2003} is coupled to a set of
classical rate equations describing the current dynamics of
ChR2.\cite{nikolic_photocycles_2009,evans_pyrho_2016}  The
multi-timescales model is seen to reproduce the chirp dependence and
other qualitative features of the experimental phase control of ChR2.

\section{Theory}

\subsection{Channelrhodopsin-2}
Channelrhodopsins have been extensively studied for their use in
optogenetics.\cite{lorenz-fonfria_channelrhodopsin_2014}  An accurate
rate model for the millisecond timescale polarization of neuron cells
expressing ChR2 has been previously
devised,\cite{nikolic_photocycles_2009,
  grossman_modeling_2011,foutz_theoretical_2012,
  evans_pyrho_2016}
and will be helpful in our analyzing the demonstrated phase control of
current.\cite{paul_coherent_2017}  Note that the focus here must be on
qualitative features of the measured phase-dependent current since a
quantitative analysis over all relevant timescales is beyond the state
of the art.

\begin{figure}[h]
  \includegraphics{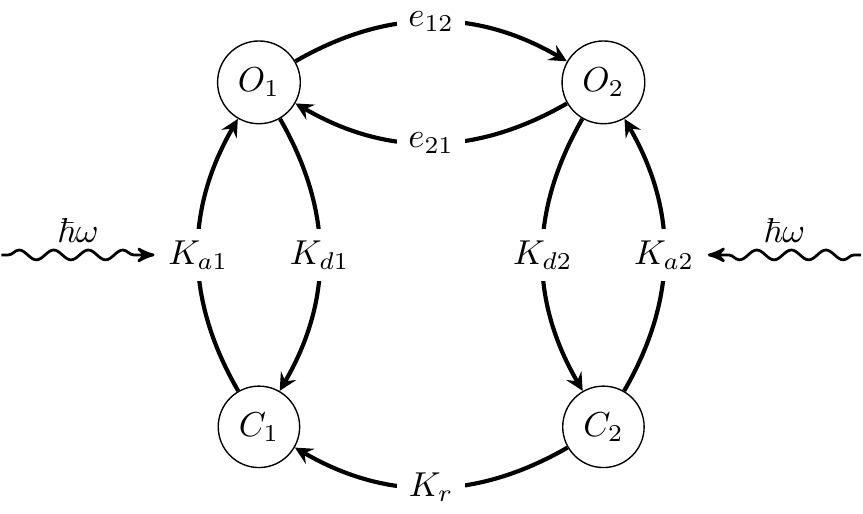}
  \caption{
    Four state model of ChR2 from
    Ref.~\onlinecite{nikolic_photocycles_2009}. $C_1$ and $O_1$ denotes the
    non-conductive (i.e., closed) and conductive (open) configurations of
    the protein in its dark-adapted form, with corresponding light-adapted
    configurations $C_2$ and $O_2$.  Arrows denote transitions between
    these states with corresponding rates (see Appendix
    \ref{sec:model-current}).}
  \label{fig:rate_model}
\end{figure}

The ChR2 protein is thought to exist in four states (see
Fig.~\ref{fig:rate_model}), with distinct spectroscopic and
electrochemical
properties.\cite{nikolic_photocycles_2009,schneider_biophysics_2015}
The protein contains a retinal molecule, which acts as the
photoreceptor.  The dark-adapted state $C_1$ is closed to ion transfer
into the cell and thus non-conductive.  Light-mediated activation,
through retinal photoisomerization, leads to the open conductive state
$O_1$ (with rate $K_{a1}$) which exists in equilibrium with the more
stable yet less conductive light-adapted open state $O_2$.  The
forward and backward transitions between $O_1$ and $O_2$ are primarily
thermally driven, with a weak dependence on the applied light.  ChR2
in the $O_1$ state can relax thermally back to the closed state $C_1$,
while ChR2 in the $O_2$ state relaxes to the light-adapted state
$C_2$.  The state $C_2$ can be activated back to $O_2$ (with rate
$K_{a2}$), also through a photoisomerization.  In the absence of
radiation however, the protein in the light-adapted state $C_2$ slowly
relax back to the dark-adapted state $C_1$.  These processes are shown
in Fig.~\ref{fig:rate_model}.

The activation rates $K_{a1}$ and $K_{a2}$, i.e. the rates of the $C_1
\rightarrow O_1$ and $C_2 \rightarrow O_2$ transitions, are given by
the photoisomerization rate of retinal in ChR2.  Importantly, the
coherent control of peak current described below results from
controlling these activation rates.  The rate of photoisomerization is
a function of the amount of light received by retinal (i.e. the number
of excitations per second) and the quantum yield of isomerization
(i.e. the fraction of excitations that goes on to form the final
product.) The activation rate for a weak, monochromatic light source
with frequency $\omega$ is given by the standard photochemical
expression,
\begin{align}
  K_{ai}(\omega) = \Phi(\omega) \sigma_\text{ret}(\omega) \eta_i(\omega) \label{eq:inco_activation_rate}
\end{align}
where $\eta_i(\omega)$ is the quantum yield of isomerization,
$\sigma_\text{ret}(\omega)$ is the cross-section of retinal,
$\Phi(\omega)$ is the spectral photon flux, and the cross-section has
been taken to be identical for both the dark- and light-adapted
states.\cite{grossman_modeling_2011}  However, this activation rate
formula is not valid in the case of two-photon excitation using a
pulsed laser; in particular, the one-photon, monochromatic activation
rate is not phase-dependent and thus is not phase controllable.  In
contrast, the treatment below, that generalizes
the activation rate formula to two-photon excitation by trains of
ultrashort pulses, is phase-dependent and phase-controllable.

The developments below focus on the coherent control of the peak
current resulting from the formation of the dark-adapted open state
$O_1$.  ChR2 current dynamics obtained using the model of Appendix
\ref{sec:model-current} is demonstrated in Fig.~\ref{fig:inco} and is obtained via
a two-step process.  Initially, the current shows a peak which
is primarily determined by the rate of formation of $O_1$ due to
excitation from
$C_1$.\cite{rickgauer_two-photon_2009,nikolic_photocycles_2009}  The
peak current is followed by a decay to a steady-state current, with a
value determined by the competing rates of formation, decay and
interconversion of the dark- and light-adapted open states.  At
moderate to high laser intensity, control of the peak
current\cite{paul_coherent_2017} is primarily a result of controlling
the initial generation of $O_1$, which is well-known to be triggered
by retinal isomerization within the
protein.\cite{rickgauer_two-photon_2009}  This latter feature allows us
to address the nature of phase control in this system.

\begin{figure}[h]
  \includegraphics[width=\textwidth]{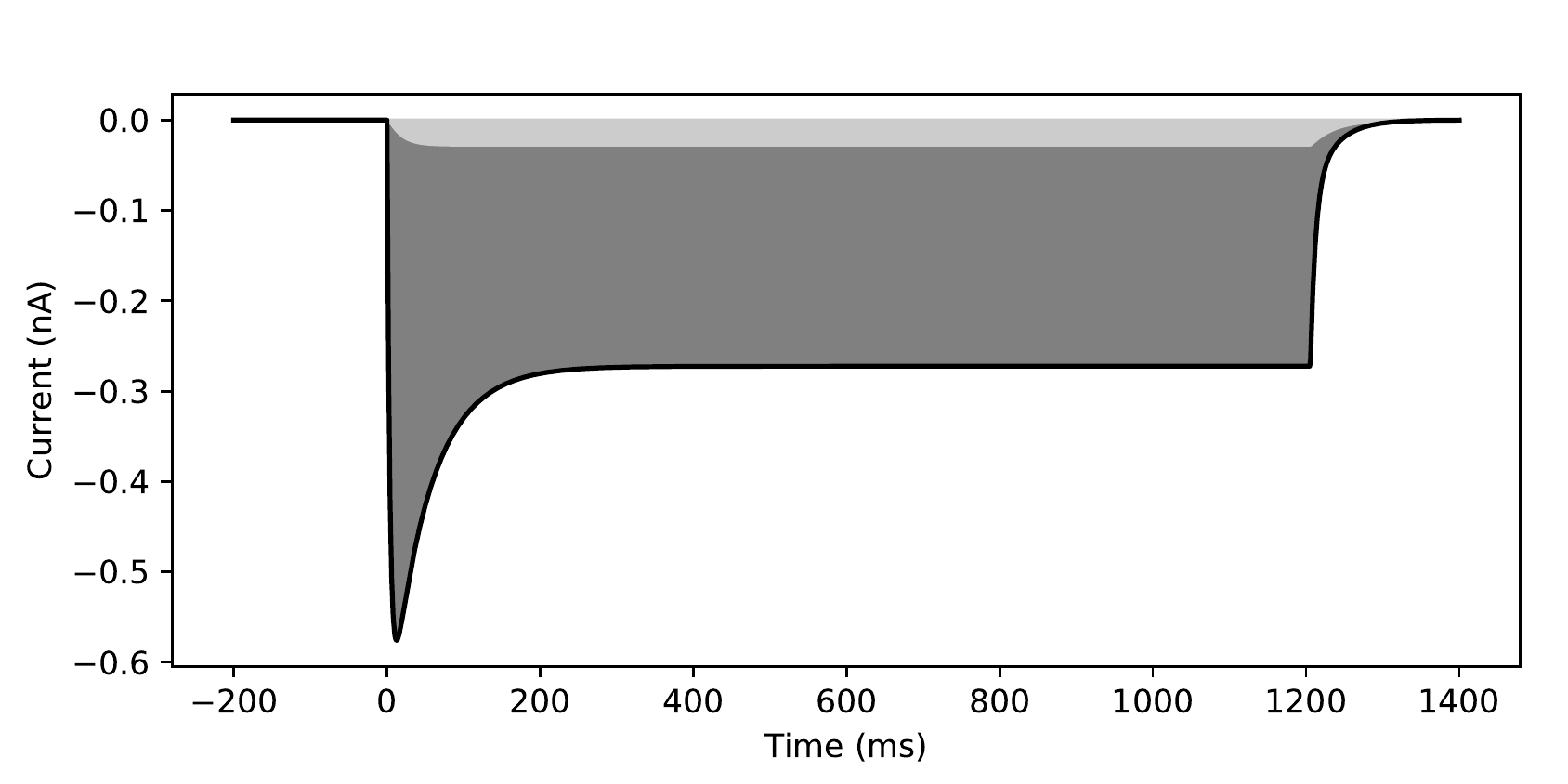}
  \caption{
    Current computed from the ChR2 conductance model described
    in Appendix \ref{sec:model-current} for a moderately intense irradiance
    (1 mW/mm$^2$), visible spectrum (470 nm), incoherent light source.  The
    excitation rate of retinal (190 photons per second using the
    cross-section $\sigma_\text{ret} = 8\times 10^{-8} \mu $m$^{-2}$) is
    comparable to that obtained by two-photon excitation in
    Ref.~\onlinecite{paul_coherent_2017} and simulations below.  Current
    contributions from the dark- and light-adapted open states of ChR2 are
    represented by the dark and light grey regions.
  }
  \label{fig:inco}
\end{figure}

\subsection{Timescales of excitation, control and measurement}
Macroscopic coherent control, as observed in
Ref.~\onlinecite{paul_coherent_2017}, operates on multiple timescales
spanning more than fifteen orders of magnitude.  Careful study of the
dynamical processes at play uncovers, as shown in this section, a
hierarchy of separated timescales.  Significantly, exploiting these
separations is key to understand how control of millisecond dynamics
can arise from ultrashort interactions.

Consider the coherent control of a photochemical reaction,
e.g. the photoisomerization of retinal, that has a metastable,
macroscopically long-lived product with decay rate $k_p$.  Excitation
is performed by a pulsed laser with a field of the following form,
\begin{align}
  \field(t) = \sum_{n=0}^{n_\text{pulses}} \field_1(t - n\Delta t)\label{eq:train}
\end{align}
where $n_\text{pulses}$ is the total number of pulses in the pulse
train with a repetition rate $k_\text{rep}=1/\Delta t$, and
$\field_1(t)$ is the field of a single pulse.  The pulses consists of a
slow envelope $A(t)$ with duration $\sigma_t$ and a carrier frequency
$\omega_c$,
\begin{align}
  \field_1(t) = A(t) \sin(\omega_c t).\label{eq:slowenvt}
\end{align}
The single pulse $\field_1(t)$ generates coherent dynamics in the
molecular system.  These dynamics decay with a decoherence time
$\tau_\text{deco} = 1/k_\text{deco}$.

Here, two-photon phase control is studied in the case where a strict
separation of timescales holds.  First, the duration of the pulse
$\sigma_t$ is much slower then the period of the carrier frequency
$\omega_c$.  Pulses are well separated compared to coherent timescales
of the system and the field, i.e. the time between pulses
($1/k_\text{rep}$) is much longer than $\sigma_t$ and
$\tau_\text{deco}$.  The macroscopic dynamics of the system (with a
characteristic time of $1/k_p$) are themselves much slower than the
laser pulse train.  These temporal relationships can be summarized as
follows,
\begin{align}
  1/\omega_c \ll (\sigma_t \text{ and } \tau_\text{deco}) \ll 1/k_\text{rep} \ll 1/k_p.\label{eq:timescales}
\end{align}
This is the case, for example, in
Ref.~\onlinecite{paul_coherent_2017}, where macroscopic control is the
result of processes occurring over fifteen orders of magnitude in
time: excitation consists of a train of near-infrared ($1/\omega_c <
1$ fs), ultrashort pulses ($\sigma_t\approx 0.1 - 1$ ps).  The pulses
induce ps timescale dynamics, which decay within $1/k_\text{deco} \ll 1 $
ns.\cite{verhoefen_photocycle_2010,neumann-verhoefen_ultrafast_2013}
The train has an 80 MHz repetition rate ($1/k_\text{rep}=16 $ ns).
Overall control is achieved over the macroscopic current dynamics of a
neuron, which are the result of large-scale conformational changes in
ChR2, with timescales $1/k_p\sim 1-1000 $ ms.  Important
simplifications can be made based on each of the above relationships
to obtain physically motivated coherent control mechanisms for the
experiment in Ref.~\onlinecite{paul_coherent_2017}.  A similar
analysis is possible for other experimental scenarios where long-lived
photoproducts are obtained from excitation with repeated ultrashort
pulses.

\subsection{Perturbative description of two-photon processes}
Two-photon absorption results from the interaction between a quantum
mechanical system and a time-dependent electric field at an order
quadratic in the intensity of the field.  Here, a semiclassical,
perturbative description of the light-matter interaction is employed.

Consider a molecule interacting with a time-varying electric field in
the dipole approxi\-mation.\cite{mukamel_principles_1995,
shapiro_quantum_2012} The Hamiltonian for such a system is given by,
\begin{align}
  H(t) = H_0 - \field(t) \mu.
\end{align}
where $H_0$, the material Hamiltonian, includes everything but the
field, i.e., the molecule and its environment, $\mu$ is the dipole
operator and $\field(t)$ is the electric field at $\vec r_0$, the
position of the molecule.  The corresponding Liouville equation for the
density matrix $\rho(t)$ is
\begin{align}
  \ddt{\rho(t)} = \super L_0\rho(t) + \frac{i}{\hbar}\field(t) \super V\rho(t), \label{eq:lvneq}
\end{align}
with the Liouvillian superoperators $\super L_0 \rho =
\commute{H_0}{\rho}/i\hbar$ and light-matter coupling superoperator
$\super V \rho = \commute{\mu}{\rho}$.  Provided that the field
vanishes sufficiently quickly, the one-photon and two-photon
contributions to $\rho(t)$ (derived in Appendix \ref{sec:pertub}) are
given by
\begin{align}
  \rho_2(t) &= \left(\frac{i}{2\pi\hbar}\right)^2  \iint_{-\infty}^{\infty}\mathrm d \omega_2\mathrm d \omega_1 e^{i(\omega_2 + \omega_1) t + \eta
  t}\field(\omega_2) \field(\omega_1-i\eta) \label{eq:rho_pt2}\\
  &\quad \times \super G_0(\omega_2 + \omega_1 -i\eta) \super V \super G_0(\omega_1 - i\eta) \super V \rho_0\nonumber\\
  \rho_4(t) &= \left(\frac{i}{2\pi\hbar}\right)^4  \iiiint_{-\infty}^{\infty}\mathrm d \omega_4 \mathrm d \omega_3 \mathrm d \omega_2\mathrm d \omega_1
  e^{i(\omega_4 + \omega_3 + \omega_2 + \omega_1) t + \eta t}\field(\omega_4) \field(\omega_3) \field(\omega_2) \field(\omega_1-i\eta)\nonumber\\
          &\quad \times\super G_0(\omega_4 + \omega_3 + \omega_2 + \omega_1 -i\eta) \super V \super G_0(\omega_3 + \omega_2 + \omega_1
          -i\eta)\label{eq:rho_pt4}\\
  &\quad \times \super V \super G_0(\omega_2 + \omega_1 -i\eta) \super V \super G_0(\omega_1 - i\eta) \super V \rho_0\nonumber
\end{align}
with the Fourier-domain field and Green's function given by
\begin{align}
  \field(\omega) &= \int_{-\infty}^{\infty}\mathrm d\omega e^{-i\omega t} \field(t)\\
  \super G_0(\omega -i \eta) &= \frac{1}{i\omega - \super L_0 + \eta}
\end{align}
The subscripts 2 and 4 in Eq.~(\ref{eq:rho_pt2}) and
Eq.~(\ref{eq:rho_pt4}) denote the order of the perturbative expansion,
which is quadratic in the field amplitude for one-photon processes and
quartic in the field amplitude for two-photon processes.

The interaction between the system and radiation yields a frequency
dependent measurable change $I_\text{diff}(\omega)$ in the intensity
of the laser pulse,
\begin{align}
  I_\text{diff}(\omega) = I_\text{in}(\omega) - I_\text{out}(\omega) \label{eq:Inet}
\end{align}
where $I_\text{in}(\omega)$ and $I_\text{out}(\omega)$ are the
intensity of radiation before and after an interaction with the
sample.  This value can be computed from the microscopic treatment
above provided the sample is highly transparent, as described in
Appendix \ref{sec:pertub}.  In this case, the change in the intensity
of the light $I_\text{diff}(\omega)$ as measured by a
spectrophotometer is proportional to the spectral number of absorbed
photons $N_\Delta(\omega)$ computed for a single interacting molecule.
The dimensionless total number of absorbed photons is used below in
quantum yield calculation, and is given by
\begin{align}
 N_\Delta = \int_{-\infty}^{\infty} \mathrm d \omega N_\Delta(\omega).\label{eq:Ndelta}
\end{align}

\subsection{Two-photon phase control with an ultrashort pulse} \label{sec:slowenv}
Coherent control of long-lived products arises from interference
between multiple indistinguishable light-induced pathways, which is
the essence of coherent control.\cite{shapiro_quantum_2012}  A weaker
form of this statement is useful here, based on frequencies of
excitation, as discussed below.  The two-photon contribution to the
time-dependent expectation value of an operator $B(t) = \Tr B
\rho_4(t)$, e.g. the isomer population in a photoisomerization
process, can be computed from Eq.~(\ref{eq:rho_pt4}).  The integrand
has the following oscillatory time-dependence,
\begin{align}
  e^{i(\omega_1 + \omega_2 + \omega_3 + \omega_4)t}\field(\omega_1) \field(\omega_2) \field(\omega_3) \field(\omega_4)
\end{align}
Thus, long-lived control requires that the sum of frequencies
($\omega_4+\omega_3 +\omega_2 + \omega_1$) be close to zero to avoid
cancellation via the oscillatory contribution of the exponential.\cite{lavigne_interfering_2017}
The limited bandwidth of typical ultrashort lasers restricts  the
number of possible control schemes that satisfies these conditions.

The Fourier transform of the field [Eq.~\eqref{eq:slowenvt}] is given
by
\begin{align}
  \field(\omega) = A(\omega-\omega_c) + A^*(\omega + \omega_c)\label{eq:slowenv}
\end{align}
where the width of $A(\omega)$ is much narrower than $\omega_c$.  Thus,
both $\field(2\omega_c)$ and $\field(\omega_c/2)$ are negligible,
which reduces the number of possible light-induced interference
processes.  Indeed, interference is only possible between pathways with
the same net number of transitions, e.g., two different two-photon
absorption pathways.  Other fourth-order processes, such as
interference between one-photon and three-photon absorption pathways,
lead to fast oscillatory contributions in
Eq.~(\ref{eq:rho_pt4}), with frequencies comparable to $\omega_c$ or a
multiple thereof (i.e. electronic coherent dynamics).

\begin{figure}[h]
  \centering
  \begin{subfigure}[b]{0.3\textwidth}
    \includegraphics{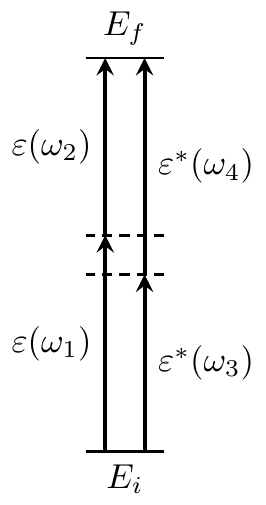}
    \caption{2 vs. 2 photon control\label{fig:scheme_2v2}}
  \end{subfigure}
  \begin{subfigure}[b]{0.3\textwidth}
    \includegraphics{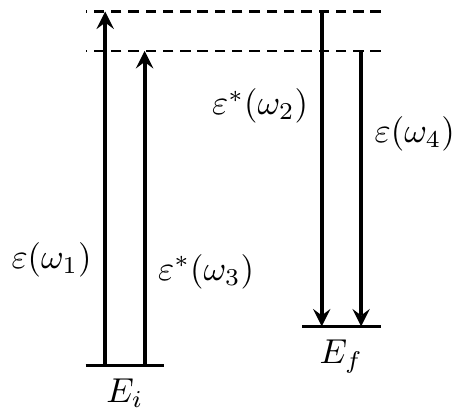}
    \caption{Pump-dump control\label{fig:scheme_pd}}
  \end{subfigure}
  \begin{subfigure}[b]{0.3\textwidth}
    \includegraphics{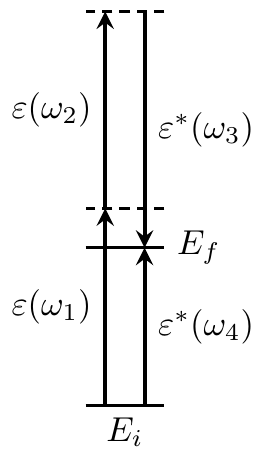}
    \caption{1 vs. 3 photon control\label{fig:scheme_1v3}}
  \end{subfigure}
  \caption{
    Three proposed coherent control schemes for stationary
    phase control in a two-photon pulsed experiment.  In each case, the
    transition probability between initial states with energy $E_i$ and
    final states with energy $E_f$ is a function of the amplitude and
    phase of the electric field $\field(\omega)$ at four distinct
    frequencies.\label{fig:allschemes}}
\end{figure}

Three control mechanisms can be proposed which account for all
possible two-photon interfering pathways respecting the above
conditions, and are shown in Fig.~\ref{fig:allschemes}.  Here $E_i$
and $E_f$ denote the energy of the initial and final states; the
overall energy imparted by the field $\field(\omega)$ is $\Delta E =
E_f -E_i$, which is approximately $2 \hbar \omega_c$ for two-photon
absorption, zero for a pump-dump process and $\hbar \omega_c$ for 1
vs. 3 photon control.

The three control mechanisms of Fig.~\ref{fig:allschemes} are best
understood in the framework of coherent
control.\cite{shapiro_quantum_2012,lavigne_interfering_2017}  Indeed,
each control mechanism is also a coherent control scheme --- a
physically motivated, rationally-designed procedure to control an
observable through interference between light-induced pathways.

Consider \textit{2 vs. 2 photon control} (Fig.~\ref{fig:scheme_2v2})
where control is the result of interference between a two-photon
absorption pathway with frequencies $\omega_1$ and $\omega_2$ and a
different two-photon absorption pathway with frequencies $\omega_3$
and $\omega_4$.  Coherent excitation through both pathways creates a
superposition with the following qualitative form,
\begin{align}
  \ket{\Psi} = \field(\omega_1) \field(\omega_2) \ket{\alpha} + \field(\omega_3) \field(\omega_4) \ket{\beta}
\end{align}
where the two terms in the RHS are the wavefunctions produced by
either two-photon absorption pathways.  Then, the expectation value of
an operator $B(t)$ resulting from $\ket{\Psi}$, e.g. the
isomerization probability of an excited molecule, is given by,
\begin{align}
  \braket{\Psi|B|\Psi} &= |\field(\omega_1)|^2 |\field(\omega_2)|^2 \braket{\alpha|B|\alpha} + |\field(\omega_3)|^2 |\field(\omega_4)|^2 \braket{\beta|B|\beta}\\
  &\quad + 2 \text{Re}\left[\field(-\omega_1) \field(-\omega_2) \field(\omega_3) \field(\omega_4) \braket{a| B| b}\right]\nonumber
\end{align}
where the Hermitian property $\field(-\omega) =\field^*(\omega)$ has
been used.  Only the last term, an interference contribution, is
sensitive to the phase of the exciting light.  Stationary coherent
control (i.e. where $\omega_1 + \omega_2 = \omega_3 + \omega_4$ and where the controlled populations persist after the pulse is over) of
this type is shown diagrammatically in Fig.~\ref{fig:scheme_2v2}.
Similarly, two-photon \textit{pump-dump control} is obtained through
interference between different absorption-stimulated emission
pathways, with frequencies $\omega_1, -\omega_2$ and $\omega_3,
-\omega_4$.  A diagram for stationary control of this type, i.e.,
with $\omega_1 -\omega_2 = \omega_3 - \omega_4$, is shown in
Fig.~\ref{fig:scheme_pd}.  Finally, a one-photon absorption pathway
interferes with a pump-pump-dump pathway to obtain \textit{1 vs. 3
photon control} of absorption at $\omega_c$, as shown in
Fig.~\ref{fig:scheme_1v3}.  All three control schemes are two-photon
(i.e. quadratic in the intensity of light) and coherent (i.e.
dependent on the phase of the light).

It should be noted that the diagrams in Fig.~\ref{fig:allschemes} are
simplifications; Eq.~(\ref{eq:rho_pt4}) is integrated over all
frequencies, and multiple schemes may be operating at the same time.
Furthermore, the initial steady state may be a mixture of energy
eigenstates, as is the case if, for example, the system is initially
thermally distributed.  Importantly, the final energy $E_f$ may be
multiply degenerate, allowing for control not only over the transition
probability from $E_i$ to $E_f$ but also over a particular state at
$E_f$.  This is the case analyzed below, where coherent control
modifies both the overall two-photon absorption \textit{and} the final
isomer population.

\subsection{Control amplification from repeated interactions}\label{sec:amplification}
Two-photon absorption is proportional to the squared peak intensity of
the exciting laser.\cite{xu_measurement_1996}  Hence, using a pulsed
laser with a high peak intensity, two-photon absorption is possible at
a very weak average laser intensity.  This well-known property has
important consequences for multiphoton control: although each pulse of
the laser may excite only a small fraction of the sample, a
considerable amount of phase-controllable absorption can be generated
due to the interaction with a large number of pulses.  This process of
control amplification from repeated interactions is described below,
with a focus on retinal isomerization in ChR2.

The photoactivation of ChR2 from the non-conductive $C_1$ state to the
conductive $O_1$ state is driven by retinal photoisomerization.  That
is, excitation with light triggers the isomerization of primarily
all-\textit{trans} retinal in $C_1$ ChR2 to a mixture of \textit{cis}
isomers.\footnote{
  The mechanism of ChR2 activation is quite
  complex.\cite{schneider_biophysics_2015}  For simplicity, we assume
  here that the photoisomerization of a retinal in a closed-state ChR2
  always produce the open state.}
The resultant ChR2 in the $C_1$ protein state containing an isomerized
retinal is a precursor to the $O_1$ state: over a 200 $\mu$s interval it 
forms the open $O_1$ protein state.\cite{bamann_spectral_2008}  If the
separation in time between laser pulses $\Delta t$ is much longer than
any coherent dynamics of retinal, and the field is sufficiently weak
such that multiple excitations within 200 $\mu$s are unlikely, then
the rate of formation (in product per second) of $O_1$ from $C_1$ can
be obtained directly from the product of the repetition rate of the
laser (pulses per second) and the probability of an isomerization
event occurring from an individual pulse (product per pulse).  All
these conditions are satisfied in the live brain cell experiments in
Ref.~\onlinecite{paul_coherent_2017}.  The activation rate is then
given by,
\begin{align}
  K_{a1} = k_\text{rep} \left( I_0 P^{(1)} + I_0^2 P^{(2)} \right)\label{eq:activation_rate}
\end{align}
where $I_0$ is a dimensionless scaling factor for the intensity of the
field, $k_\text{rep}$ is the laser repetition rate and $P^{(1)}$ and
$P^{(2)}$ are the one- and two-photon contributions (from a single
laser pulse) to the quasi stationary \textit{cis} population.  These
are computed perturbatively using Eq.~(\ref{eq:rho_pt2}) and
Eq.~(\ref{eq:rho_pt4}) as follows,
\begin{align}
  P^{(1)} = \Tr[P_\text{cis} \rho_2(T)] \text{ and } P^{(1)} = \Tr[P_\text{cis} \rho_4(T)]\label{eq:contribs}
\end{align}
where $P_\text{cis} $ is the projection operator for the \textit{cis}
population, $\rho_2(T)$ and $\rho_4(T)$ are the result of excitation
from the steady-state ground \textit{trans} state of retinal, and $T$
is a time much longer than ultrafast coherent dynamics.  The $P^{(2)}$
term is dependent on the laser phase and hence is the origin of the
control discussed below.

As noted above, repeated interactions can create a large amount of
multiphoton control even from weak pulses.  The activation rate from
Eq.~\eqref{eq:activation_rate} is seen to depend linearly on the
repetition rate of the laser.  Thus, the activation rate $K_{a1}$ can be increased by increasing
the repetition rate without affecting the relative contributions from
one, two or higher order processes.  That is, if the effect produced by
a weak pulse is highly phase dependent, the magnitude of control can
be made large simply by increasing the repetition rate, up to limits
set by the fast (relative to the ms time scale of the induced current)
dynamics of ChR2.

A comparison with CW radiation is instructive.  A CW field and a pulsed
field with the same average power $W$ will deliver, over a time
$t_\text{on}$, the same average energy $\braket{E} = W t_\text{on}$.
However, the resulting excitation dynamics are not the same.  The $N$
photon contribution to the population depends on the $N$-th power of
overall field intensity before $t_\text{on}$,
\begin{align}
  P_\text{CW}^{(N)} \propto \left(t_\text{on} I_0\right)^{N}.
\end{align}
In general, the proportion of each order of the perturbative expansion
changes with the intensity of the field \textit{and} the length of
irradiation.  That is not the case for the pulsed field above, where
only the intensity of individual pulses affect the relative
proportions of multiphoton processes,
\begin{align}
  P_\text{pulsed}^{(N)} \propto k_\text{rep} t_\text{on} I_0^{N}.
\end{align}
In effect, excitation with two sufficiently separated and sufficiently
weak pulses produces twice as much product as excitation with one
pulse; absolute yields can thus be increased by increasing the number
of interactions through $t_\text{on}$ and $k_\text{rep}$ without a
concomitant increase in deleterious higher order transitions that
would accompany an increase in the overall intensity of the field or,
in the case of continuous irradiation, a longer exposure to light.\cite{xu_measurement_1996}

\section{Results and discussion}
Microscopic mechanisms and a physical description of control
amplification were proposed above to describe large phase control
effects arising from repeated two-photon excitation with ultrafast
pulses, of the kind demonstrated in a recent experiment on
ChR2-expressing neurons.\cite{paul_coherent_2017}  In the first
section below, the physics underlying each of the three proposed
coherent control schemes is examined using computationally simple
models.  In the second section, qualitative experimental features are
reproduced using a quantum model of retinal
dynamics\cite{hahn_quantum-mechanical_2000,
hahn_ultrafast_2002,balzer_mechanism_2003} coupled to realistic
classical rate equations for ChR2
photocurrents.\cite{nikolic_photocycles_2009,
grossman_modeling_2011,evans_pyrho_2016}

\subsection{Mechanisms of two-photon coherent control}\label{sec:mech-two-phot}
In this section, the three two-photon coherent control mechanisms
proposed above are studied using variations of a simple model.
Parameters are chosen so as to isolate each of the three control
schemes.

\begin{figure}[h]
  \centering
  \begin{subfigure}[b]{0.4\textwidth}
    \includegraphics{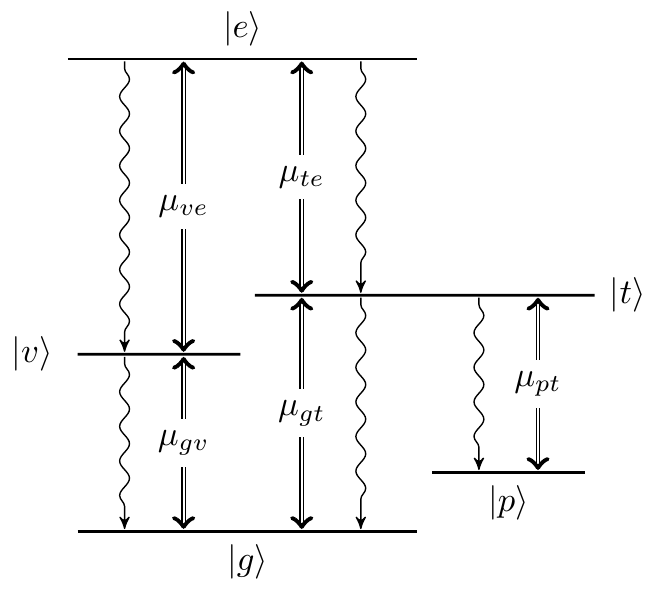}
    \caption{\label{fig:model}}
  \end{subfigure}
  \begin{subfigure}[b]{0.4\textwidth}
    \includegraphics{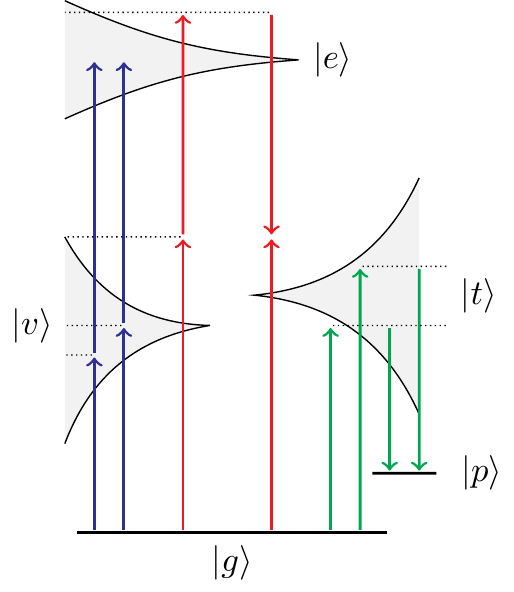}
    \caption{\label{fig:model_control}}
  \end{subfigure}
  \caption{
    (a) System used to characterize each of the three control
    schemes described above.  Solid arrows denote light-induced excitation
    pathways while curved arrows denote bath-induced relaxation pathways.
    The states $\ket{v}$, $\ket{t}$ and $\ket{e}$ are metastable
    resonances with finite lineshapes. (b) Excitation pathways for 2 vs. 2
    control (blue), 1 vs. 3 control (red) and pump-dump control (green)
    through the $\ket{v}$ resonance, the $\ket{v}$ and $\ket{e}$
    resonances and the $\ket{t}$ resonance.}
  \label{fig:models}
\end{figure}

A model for each of the three control scheme is constructed by tuning
the parameters of the minimal five-level system shown in
Fig.~\ref{fig:model}.  The system has two steady states $\ket{g}$ and
$\ket{p}$, the initially populated ground state and the controlled
product state.  The other three states of the system are unstable
\textit{resonances}, due to relaxation processes terminating at either
$\ket{g}$ or $\ket{e}$.  Relaxation and decoherence are treated in the
Lindblad formalism\cite{weiss_quantum_2012} parametrized by rates.  By
tuning the dipole transition elements $\mu_{ij}$ and the system
energies $E_i$, each of the three control schemes can be separately
selected, as described below. Parameters that expose each model as
well as Lindblad rates are provided in Appendix
\ref{sec:minimal-models}.

\textit{Control by a 2 vs. 2 scheme}, shown in blue in
Fig.~\ref{fig:model_control}, is obtained through the interference of
distinct (i.e. with distinct frequencies $\omega_1 + \omega_2$ and
$\omega_3 +\omega_4$) two-photon pathways in the presence of an
intermediate resonance $\ket{v}$.  The two-photon absorption
cross-section to $\ket{e}$ is then proportional to the product of a
one-photon excitation to $\ket{v}$ followed by a one-photon excitation
to $\ket{e}$,
\begin{align}
   T^{(2)}_{g \rightarrow e} \propto I_0^2 |\mu_{g,v} \mu_{v,e}|^2
\end{align}
where $I_0$ is the intensity of the exciting pulse.  If the $\ket v
\rightarrow \ket e$ transition is significantly brighter than the
$\ket g \rightarrow \ket v$ transition and the $\ket{v}$ state is
near-resonant, considerable two-photon absorption is obtained without
saturating the one-photon $\ket{g}\rightarrow\ket{v}$ transition.  A
model for 2 vs. 2 control is obtained here by choosing $\mu_{ve} = 100
\mu_{gv}$ and $E_v\approx E_e/2$.  Such a model can represent, for
example, two-photon near-infrared absorption through a highly excited
nearly dark vibrational state.

The \textit{pump-dump control scheme}, shown in green in
Fig.~\ref{fig:model_control}, is the result of interference between
distinct pump-dump pathways in a manner similar to the 2 vs. 2 control
scheme above.  Pump-dump control also requires an intermediate
resonance (the transition state $\ket{t}$), but excitation to the
intermediate resonance is followed by stimulated emission to the final
state $\ket{p}$, as opposed to further excitation in the 2 vs. 2
control scheme.  This control scheme requires nonzero $\mu_{g,t}$ and
$\mu_{t, p}$, which generate a two-photon pathway from $\ket{g}$ to
$\ket{p}$ via the lineshape of the resonance $\ket{t}$.  Significant
pump-dump control requires $\ket{p}$ to be of similar energy to
$\ket{g}$, such that both pump and dump transitions are resonant,
another significant difference with the 2 vs. 2 control scheme.
Finally, as above, control without saturation is made possible by
selecting $\mu_{t,p} \gg \mu_{g,t}$, and $\ket{t}$ to be near resonant
with the near-infrared excitation pulse.

Finally, \textit{ 1 vs. 3 control} arises from interference between a
one-photon excitation pathway and a three-photon excitation and
stimulated emission pathway.  In this model study, the former is
provided by a one-photon excitation to the lineshape of $\ket{t}$
while the latter is the result of a non-resonant two-photon excitation
through the resonances $\ket{v}$ and $\ket{e}$, followed by stimulated
emission to $\ket{t}$.  Thus, the 1 vs. 3 contribution obeys the
following,
\begin{align}
  T^{(2)}_{g \rightarrow t} \propto I_0^2 |\mu_{g,v}\mu_{v,e}\mu_{e, t} \mu_{g, t}|.
\end{align}
The 1 vs. 3 contribution depends linearly on the $g\rightarrow t$
transition \textit{amplitude}, whereas both the one-photon absorption
to $\ket{t}$ and two-photon absorption to $\ket e$ via the near
resonant state $\ket{t}$ depend quadratically on $\mu_{g,t}$.  Strong
control without saturation and without resonant two-photon excitation
thus requires $\mu_{g,t}$ to be very small compared to $\mu_{g,v}$,
$\mu_{v,e}$ and $\mu_{e,t}$.  Otherwise, any control will be lost to a
high one-photon absorption to $\ket{t}$ or a high non-resonant
two-photon absorption to $\ket{e}$.  A model for this system is
obtained by choosing $\ket{v}$ to be far from resonance, $\ket{t}$ to
be resonant and $\mu_{g,t}$ to be small and non-zero.

Of course, it may not be possible to cleanly separate the three
control schemes in an experiment.  Indeed, depending on the coupling
scheme and intensity of the exciting radiation, all three control
mechanisms may simultaneously contribute to coherent control.  The
approach in artificially selecting parameters that promote one or the
other control mechanism, as outlined below, is not designed to
accurately reproduce control in channelrhodopsin, but rather to
expose the similarities and differences between various control
mechanisms.  In particular, clear experimental signatures of each
mechanism by which they can be identified are obtained below, which
can be used to rigorously characterize control experimentally.

\subsubsection{Control with chirped excitations}
The key experimental result of Ref.~\onlinecite{paul_coherent_2017}
shows that control is effected by chirping the exciting pulse.
Applying a chirp $\chi$ to the slow envelope pulse $\field(\omega)$ in
Eq.~\eqref{eq:slowenv} above yields,
\begin{align}                   
  \field(\omega, \chi) = A(\omega-\omega_0) e^{-i\chi (\omega-\omega_0)^2} + c.c.
\end{align}
In the time domain and for a positive chirp, this is equivalent to
applying a frequency-dependent delay, wherein the low frequency
components of the pulse arrive before the high frequency components.  A
negative chirp yields the opposite effect.  The chirp is a pure phase
modification; hence any obtained control is entirely due to phase
effects.  All three of the above control schemes are sensitive to both
the magnitude and sign of the chirp.  The exact dependence is a
function of the spectral properties at resonance, i.e. the lineshape
and frequency of the resonances in Fig.~\ref{fig:models}.

\begin{figure}[h]
  \centering
  \begin{subfigure}[b]{0.3\textwidth}
    \includegraphics[width=\textwidth]{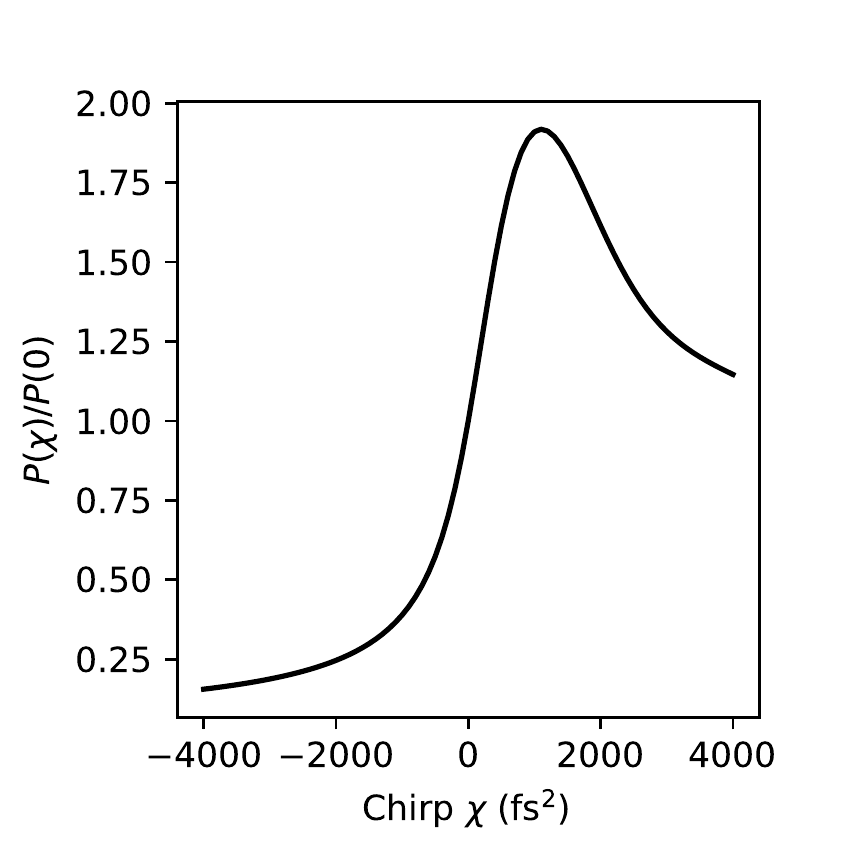}
    \caption{2 vs. 2 control}
  \end{subfigure}
  \begin{subfigure}[b]{0.3\textwidth}
    \includegraphics[width=\textwidth]{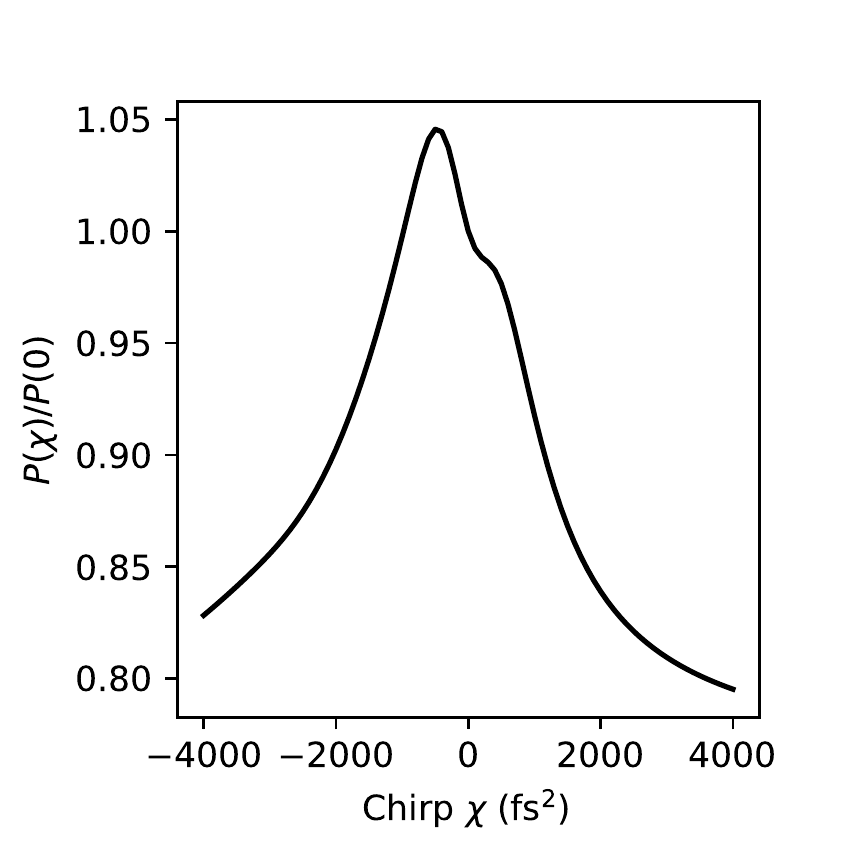}
    \caption{Pump-dump control}
  \end{subfigure}
  \begin{subfigure}[b]{0.3\textwidth}
    \includegraphics[width=\textwidth]{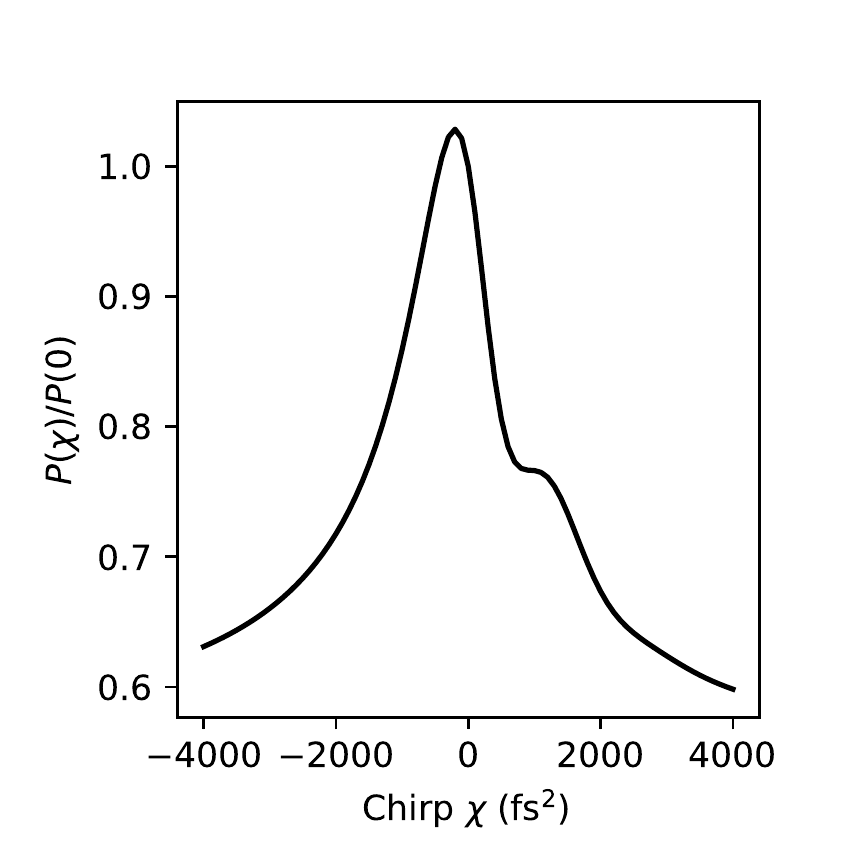}
    \caption{1 vs. 3 control}
  \end{subfigure}
  \caption{
    Probability of forming the product state $\ket{p}$
    normalized by the result at zero chirp for resonant (and hence
    controllable) models of retinal shown in Fig.~\ref{fig:models}.
  }
  \label{fig:control}
\end{figure}

The effect of chirp on the overall transition probability is shown in
Fig.~\ref{fig:control} for each of the three control schemes.  The
computed observable is the amount $P(\chi)$ of steady-state product
formed, up to fourth order in the field amplitude $\field(\omega,
\chi)$, given by
\begin{align}
  P(\chi) = \braket{p|\left(\rho_0  + \rho_2(T) + \rho_4(T)\right)|p} \label{eq:pchi}
\end{align}
where $\rho_0 = \ket{g}\bra{g}$, $\rho_2(T)$ and $\rho_4(T)$ are given
by Eq.~(\ref{eq:rho_pt2}) and Eq.~(\ref{eq:rho_pt4}), and where the
field is a near-infrared Gaussian pulse with an unchirped full width
at half maximum (FWHM) of 80 fs, a central frequency of 1.38 eV and an
applied chirp of $\chi$.  The time $T>10$ ps is well after a new steady
state is reached.

All three curves are superficially similar to one another and to that
of Ref.~\onlinecite{paul_coherent_2017}, showing that the relationship
between the excitation probability and the chirp is not sufficient to
discern the underlying control mechanism.  In all cases, response to
the sign of the chirp depends on whether the near-resonant transition
is of a lower or higher energy than the exciting pulse, and does not
provide additional information about the mechanism.  It will be shown
below, however, that certain spectral signatures can be used to
distinguish each of these control schemes.

The population $P(\chi)$ depends on the sign of the applied chirp for
all three control mechanisms (Fig.~\ref{fig:control}), as does the
experimental control of neuron current.  This is due to the presence,
in all three models, of a near-resonant one-photon transition.  A
near-resonant transition is likely to be the source of chirp control
in Ref.~\onlinecite{paul_coherent_2017} as well as nonresonant
processes are sensitive to the magnitude but not to the sign of the
chirp.\cite{meshulach_coherent_1998,lozovoy_multiphoton_2003}

\begin{figure}[h]
  \centering
  \begin{subfigure}[b]{0.2\textwidth}
    \includegraphics{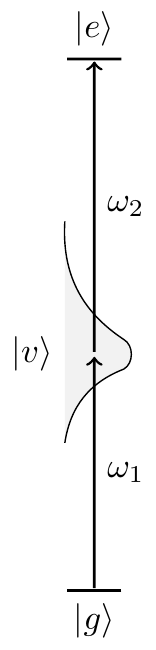}
  \caption{\label{fig:resonant}}
\end{subfigure}
  \begin{subfigure}[b]{0.2\textwidth}
     \includegraphics{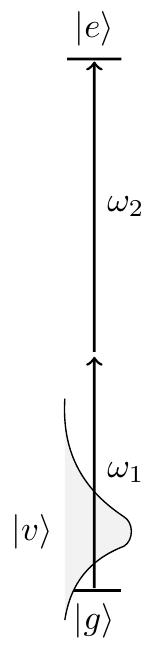}
  \caption{\label{fig:nonresonant}}
  \end{subfigure}
  \caption{
    Pictorial description of two-photon absorption through an
    intermediate state (a) at or near resonance and (b) far from
    resonance. $\omega_1$ and $\omega_2$ are two frequencies of light,
    with $\omega_1<\omega_2$.}
  \label{fig:chirp_control}
\end{figure}

The magnitude of the chirp determines the delay between high and low
frequencies of the pulse; whether low frequencies arrive before or
after high frequencies is given by the sign of the chirp.  The impact
the latter has on control is best described by an example.  Consider
the case of two-photon absorption through an intermediate state
$\ket{v}$, represented pictorially in Fig.~\ref{fig:chirp_control},
with two frequencies $\omega_1$ and $\omega_2 > \omega_1$.  The
two-photon absorption probability is maximized by exciting first to
$\ket{v}$ and then to $\ket{e}$.  If $\ket{v}$ is near-resonant with
$\omega_1$ then the transition probability will be maximized when
frequency $\omega_1$ arrives before $\omega_2$, i.e. when the chirp is
positive (Fig.~\ref{fig:resonant}). However, if $\ket{v}$ is far from
both $\omega_1$ and $\omega_2$, it matters not whether $\omega_1$ or
$\omega_2$ arrives first; any dependence on the sign of the chirp is
lost (Fig.~\ref{fig:nonresonant}).  A similar description can be made
for all three control mechanisms.

\begin{figure}[h]
  \centering
  \begin{subfigure}[b]{0.3\textwidth}
    \includegraphics[width=\textwidth]{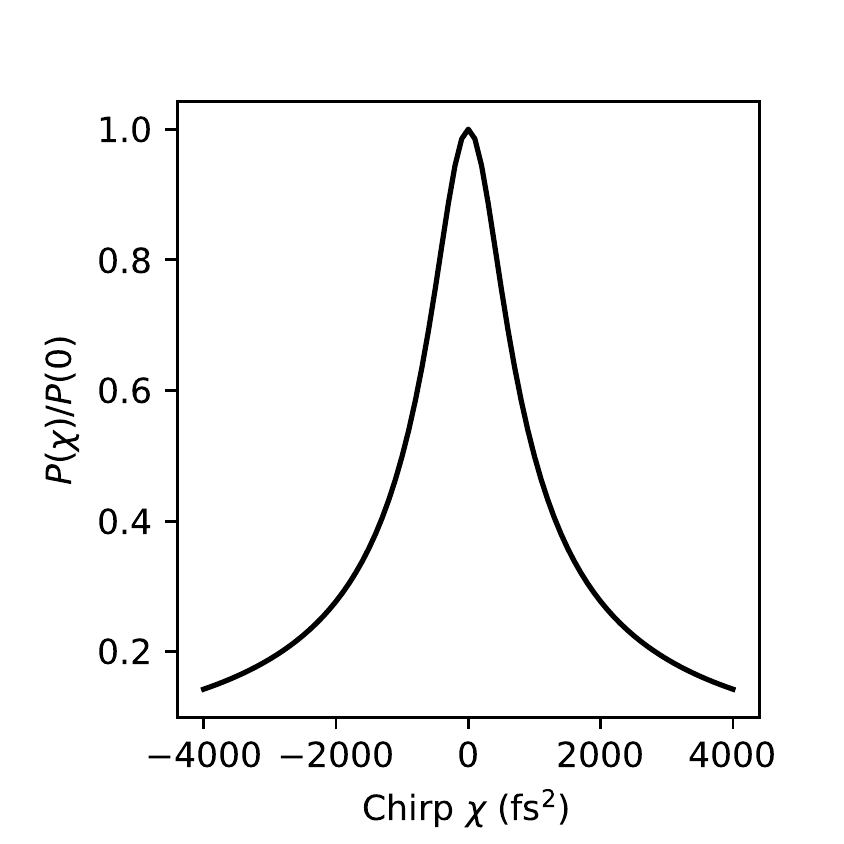}
    \caption{2 vs. 2 control}
  \end{subfigure}
  \begin{subfigure}[b]{0.3\textwidth}
    \includegraphics[width=\textwidth]{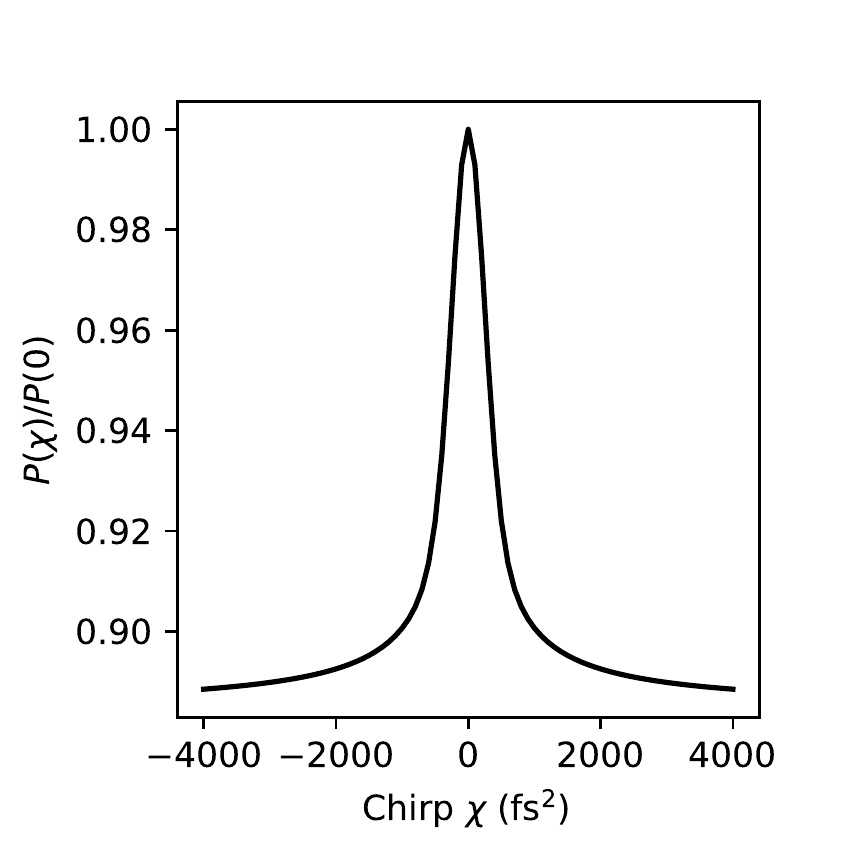}
    \caption{Pump-dump control}
  \end{subfigure}
  \begin{subfigure}[b]{0.3\textwidth}
    \includegraphics[width=\textwidth]{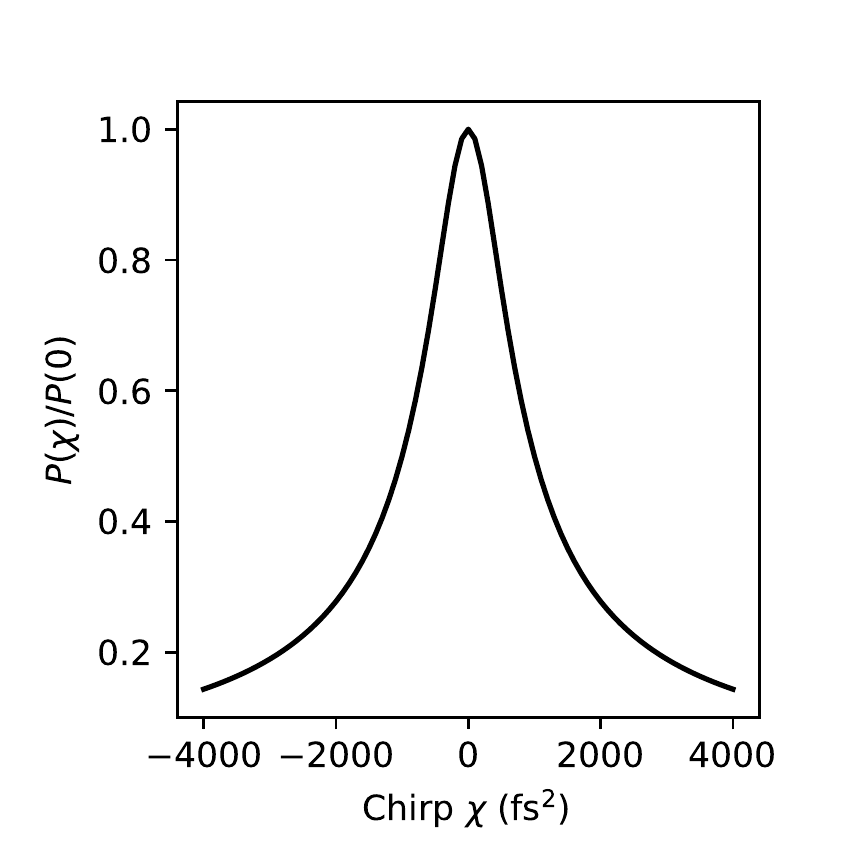}
    \caption{1 vs. 3 control}
  \end{subfigure}
  \caption{
    As in Fig.~\ref{fig:control} but without a near-resonant
    transition.  In all cases, light-induced product formation becomes
    independent of the sign of the chirp, as
    expected.\cite{meshulach_coherent_1998,ogilvie_use_2006,
      brenner_two-photon_2013}
  }
  \label{fig:nocontrol}
\end{figure}

Hence, in the absence of near-resonant transition, dependence on the
sign of the chirp is lost for all three mechanisms.  For example, in
the 2 vs. 2 case, we can examine this by lowering the system energy of
$\ket{v}$ such that the $\ket{v}$ lineshape is out of resonance.  The
same is done with $\ket{t}$ for the pump-dump.  Then, the sign of the
chirp no longer affects the transition probability, as shown in
Fig.~\ref{fig:nocontrol}.  Similarly, in the 1 vs. 3 case, in the
absence of a one-photon resonant transition to $\ket{t}$, the 1 vs. 3
contribution becomes zero.  Then, only the non-resonant two-photon
absorption pathway from $\ket g$ to $\ket e$ is bright and all dependence on the
sign of the chirp also vanishes.

\subsubsection{Absorption and quantum yield}
For all three schemes, phase control arises both from a change in
overall absorption and from a direct, phase-dependent change in the
likelihood of product formation upon absorption, i.e., the quantum
yield.  The quantum yield of a photoproduct is given by the ratio of
formed product to the number of photons absorbed.  For the present
models, the quantum $\text{QY}(\chi)$ yield is given by the following
function of the chirp $\chi$,
\begin{align}
  \text{QY}(\chi) = P(\chi) / N_\Delta(\chi)\label{eq:qy}
\end{align}
where $P(\chi)$ is defined in Eq.~(\ref{eq:pchi}) and the number of
absorbed photon is computed from Eq.~(\ref{eq:Ndelta}).

\begin{figure}[h]
  \centering
  \begin{subfigure}[b]{0.3\textwidth}
    \includegraphics[width=\textwidth]{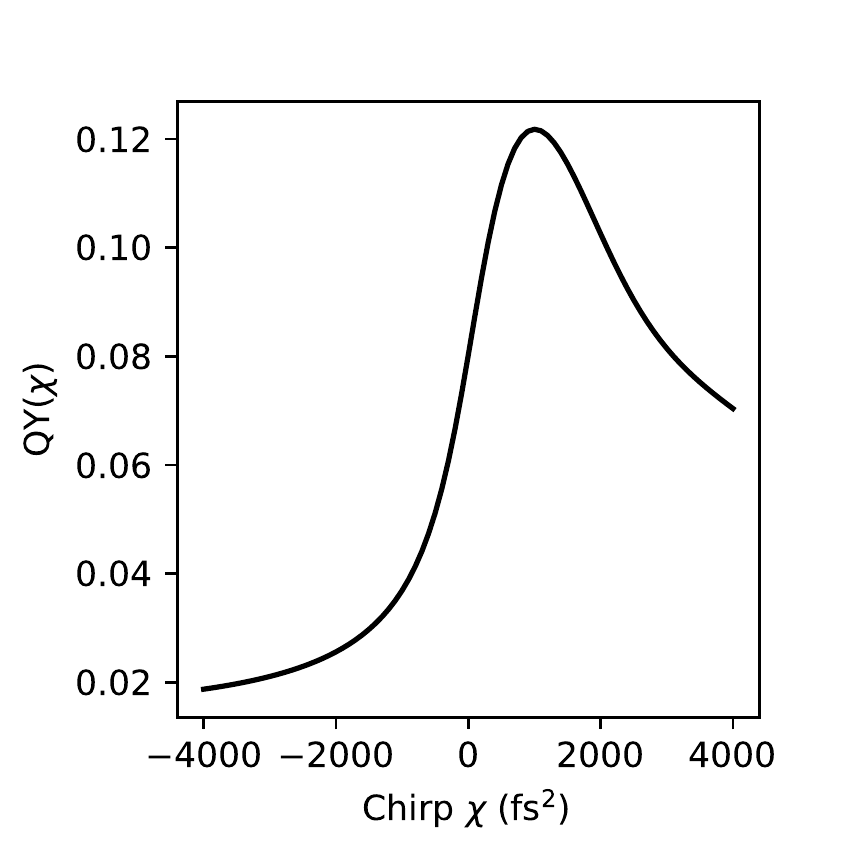}
    \caption{2 vs. 2 control}
  \end{subfigure}
  \begin{subfigure}[b]{0.3\textwidth}
    \includegraphics[width=\textwidth]{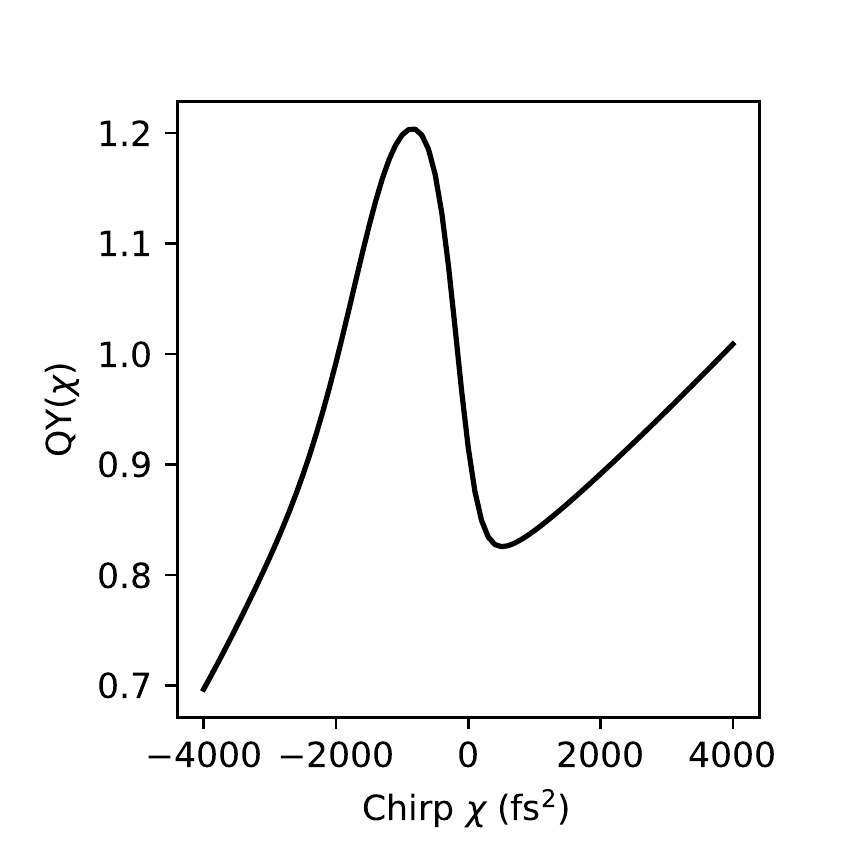}
    \caption{Pump-dump control}
  \end{subfigure}
  \begin{subfigure}[b]{0.3\textwidth}
    \includegraphics[width=\textwidth]{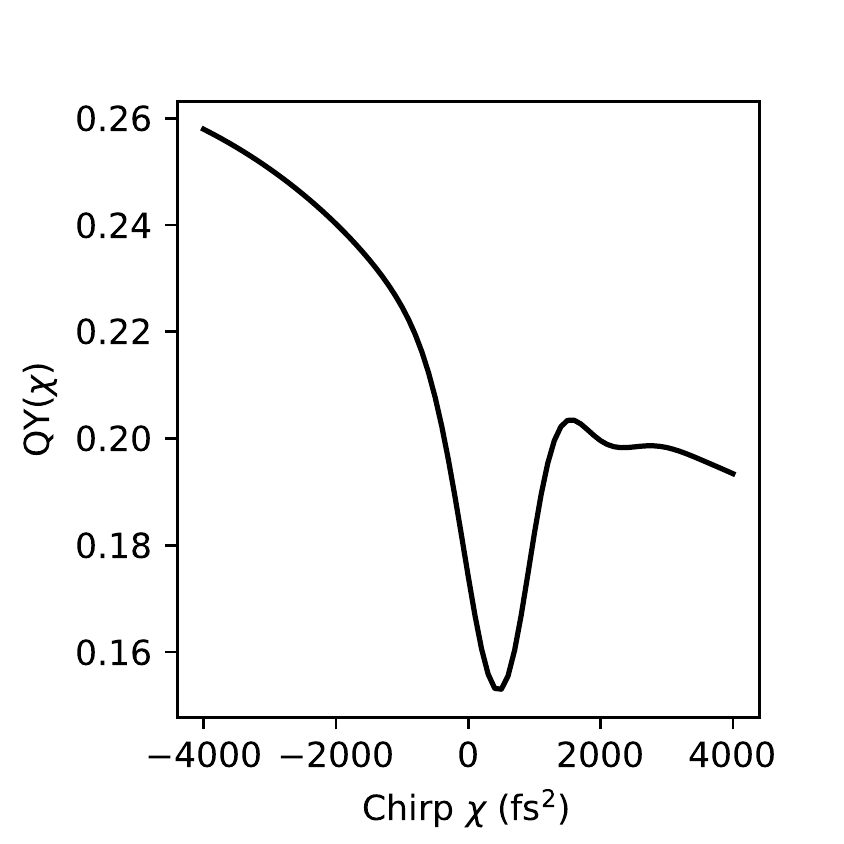}
    \caption{1 vs. 3 control}
  \end{subfigure}
  \caption{
    Quantum yield for the formation of $\ket{p}$ as a function
    of chirp computed from Eq.~(\ref{eq:qy}).}
  \label{fig:control_qy}
\end{figure}

Control over the quantum yield is demonstrated in
Fig.~\ref{fig:control_qy}. The quantum yield depends in all three
cases on the sign and magnitude of the chirp (i.e., phase controls
both the amount and the efficiency of product formation).  For the present
model, the quantum yields of Fig.~\ref{fig:control_qy} are readily
explained based on model parameters, specifically the 50\% and 80\%
yields of the $\ket{e}\rightarrow\ket{t}$ and $\ket{t} \rightarrow
\ket{p}$ transitions.  The quantum yield of the two step process from
$\ket{e}$ to $\ket{p}$ via $\ket{t}$ is 40\%, the product of the
yields of the individual steps.  Thus, the maximum quantum yield for the 2
vs. 2 control scheme in this model is 20\%; two photons are required
to form $\ket{e}$, which forms the product $\ket{p}$ 40\% of the time
(The actual value is lower, reflecting the presence of one-photon
excitations of $\ket{v}$ which does not contribute to product
formation.)  Similarly, the 1 vs. 3 control scenario contains
two-photon contributions, with a maximal yield of 20\%, as well as
direct excitation to $\ket{t}$, which has a maximal yield of 80\%. The
obtained value is significantly lower than the latter.  The quantum
yield of the pump-dump scheme is somewhat poorly defined,
\footnote{
  There is no accepted definition for the quantum yield in a pump-dump
  experiment.  The quantum yield of the pump-dump control scheme is here
  above unity as less than one photon is used per product obtained; a
  significant portion of absorbed ``pump'' photons are re-emitted by
  stimulated emission.
}
but is similarly phase controllable.

\subsubsection{Control mechanisms signatures}
The control obtained as a function of chirp is broadly similar for all
three mechanisms (as seen in Fig.~\ref{fig:control}) and thus provides
little or no information about the source of control.  However, each
of the control schemes results from a specific pattern of interactions
with the exciting field, and spectroscopic signatures that can be used
to discriminate between the three proposed control schemes are encoded
in the outgoing light from the system.  Specifically, these
interactions produce absorption and stimulated emission peaks,
spectroscopic evidence of the control mechanism, in the outgoing
radiation.  These spectroscopic signatures are discussed below.

Consider then the change in the light intensity, which for a weak
pulse is given by Eq.~(\ref{eq:Inet}).  The intensity
$I_\text{diff}(\omega)$ contains information both about the field itself
and about the system.  For linear absorption, the latter can be
isolated by computing the transmission,
\begin{align}
  T(\omega) = 1 + I_\text{diff}(\omega) / I_\text{in} (\omega).
\end{align}
However, here the intensity is used instead of the transmission, since
the latter is, in the case of multiphoton processes, no longer a
property of the material alone but also of the exciting pulse.  The
two-photon absorption cross-section is similarly affected; the
cross-section has a non-trivial dependence on the phase of the
exciting light when the excitation is ultrashort and
shaped.\cite{lozovoy_multiphoton_2003,ogilvie_use_2006}  These issues
are addressed by directly measuring $I_\text{diff}(\omega)$,
e.g. using a spectrophotometer, but without normalizing by the
spectrum of the incoming field.  The result is a function of the
incoming field but is easily and unambiguously interpreted: it is
negative at frequencies where light is absorbed and positive at
frequencies where light is emitted.

\begin{figure}
  \centering
  \begin{subfigure}[b]{0.7\textwidth}
    \includegraphics[width=\textwidth]{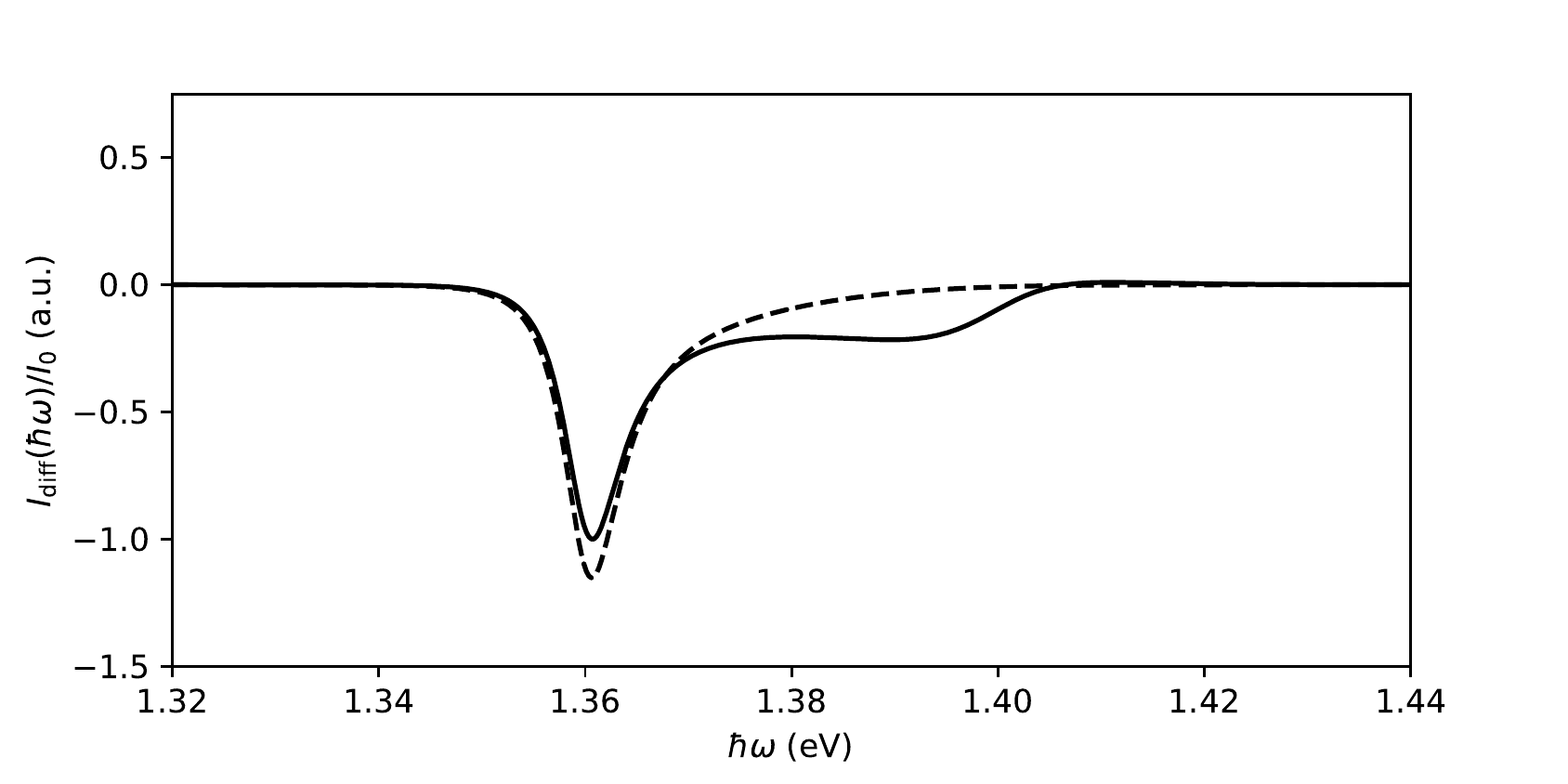}
    \caption{2 vs. 2 control}\label{fig:r2_spectra}
  \end{subfigure}
  \begin{subfigure}[b]{0.7\textwidth}
    \includegraphics[width=\textwidth]{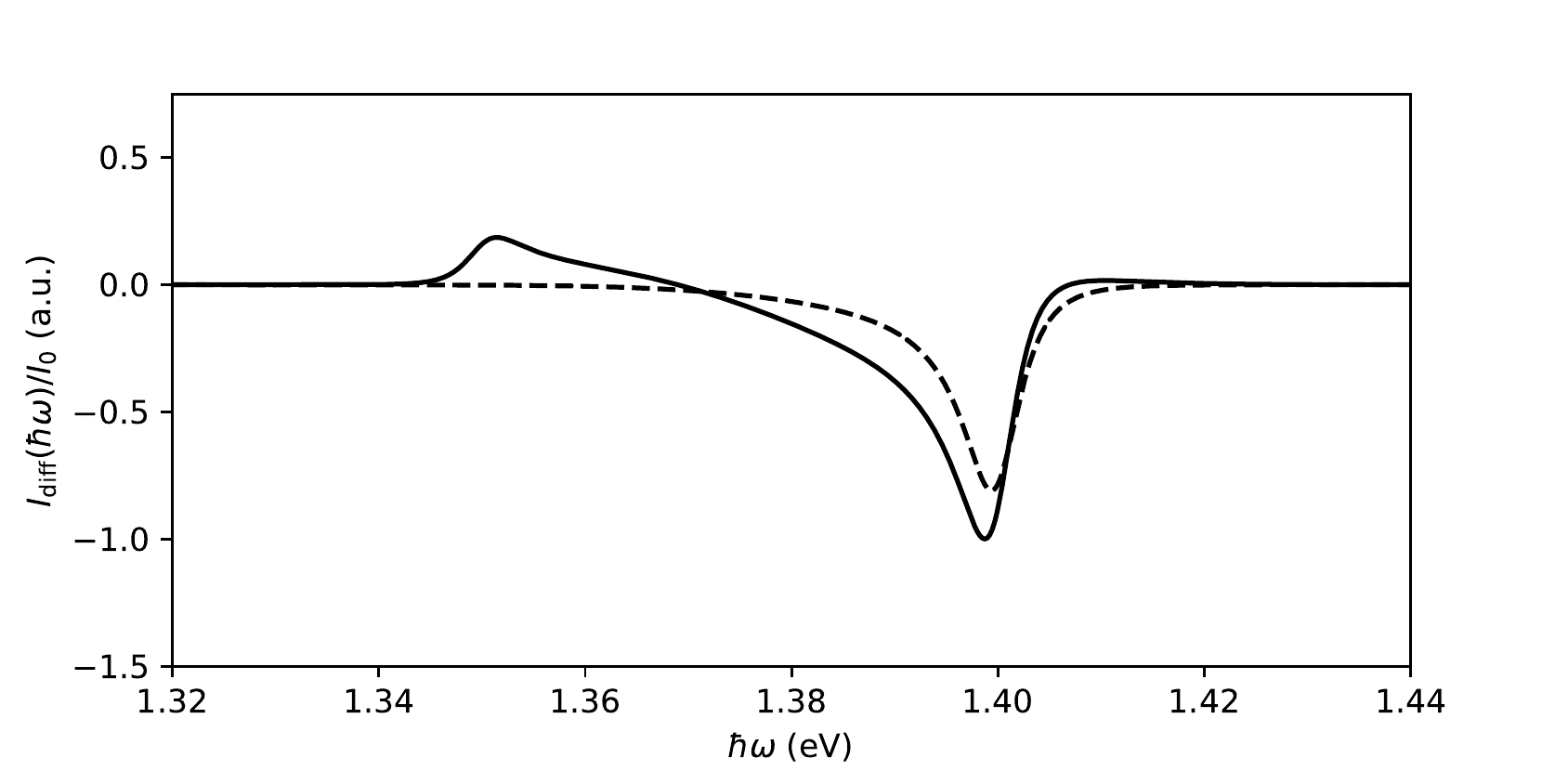}
    \caption{Pump-dump control}\label{fig:rpd_spectra}
  \end{subfigure}
  \begin{subfigure}[b]{0.7\textwidth}
    \includegraphics[width=\textwidth]{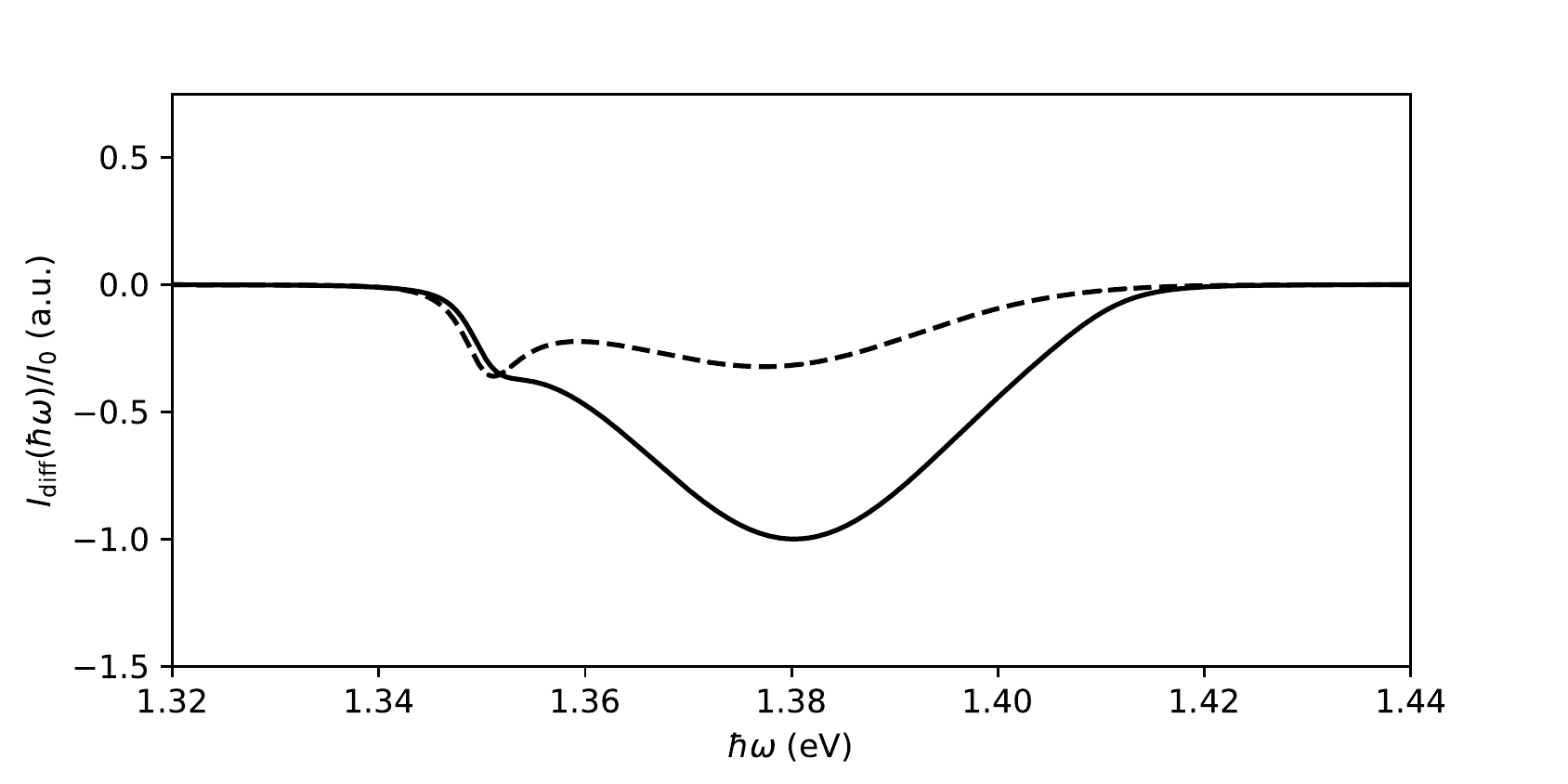}
    \caption{1 vs. 3 control}\label{fig:r3v1_spectra}
  \end{subfigure}
  \caption{
    Change in intensity $I_\text{diff}(\omega) = I_\text{in}(\omega)
    - I_\text{out}(\omega)$ for (a) 2 vs. 2, (b) pump-dump, and (c) 1 vs.
    3 control for an electric field with an intensity $I_0 = I_c$ (solid) and a
    much weaker field with intensity $I_0 = 10^{-6} I_c$.  Both are normalized by
    $I_0$.  $I_c$ corresponds to the intensity used in
    Figs.~\ref{fig:control}-\ref{fig:control_qy} at which significant
    two-photon contributions are present.}
\end{figure}

$I_\text{diff}(\omega)$ for the near-resonant 2 vs. 2 case is shown in
Fig.~\ref{fig:r2_spectra}.  The weak-field case shows absorption by a
state at $\hbar\omega_{v,g}=$1.36 eV, which is near-resonant with the
pulse centered at 1.38 eV.  The main visible transition of the system
is at $\hbar \omega_{e,g}=2.76$ eV.  At the intensity corresponding to
control shown above, two-photon absorption becomes significant.
Absorption from $\ket v$ to $\ket e$ is then seen as a two-photon
absorption dip appearing at $\hbar \omega_{e,v} = 1.40$ eV, with a
lineshape reflecting that of the field and of the excited state.
Control of this 2 vs. 2 type can thus be identified by the appearance
of an absorption peak at higher intensity that is not
present in the linear excitation regime.

The equivalent spectra for the pump-dump case is shown
Fig.~\ref{fig:rpd_spectra}.  At low intensity, only absorption at
$\hbar \omega_{g,t}=1.40$ eV is seen.  Increasing the intensity into
the control regime leads to appearance of an emissive feature at
$\omega_{t,p}=1.35$ eV, the dump frequency, where there is no
weak-field absorption.  Thus, pump-dump control has the opposite
signature as 2 vs. 2 control; it can be identified by the appearance
of an emission peak at higher intensity that is not present at lower
intensity.

\begin{figure}
  \centering
  \begin{subfigure}[b]{0.7\textwidth}
    \includegraphics[width=\textwidth]{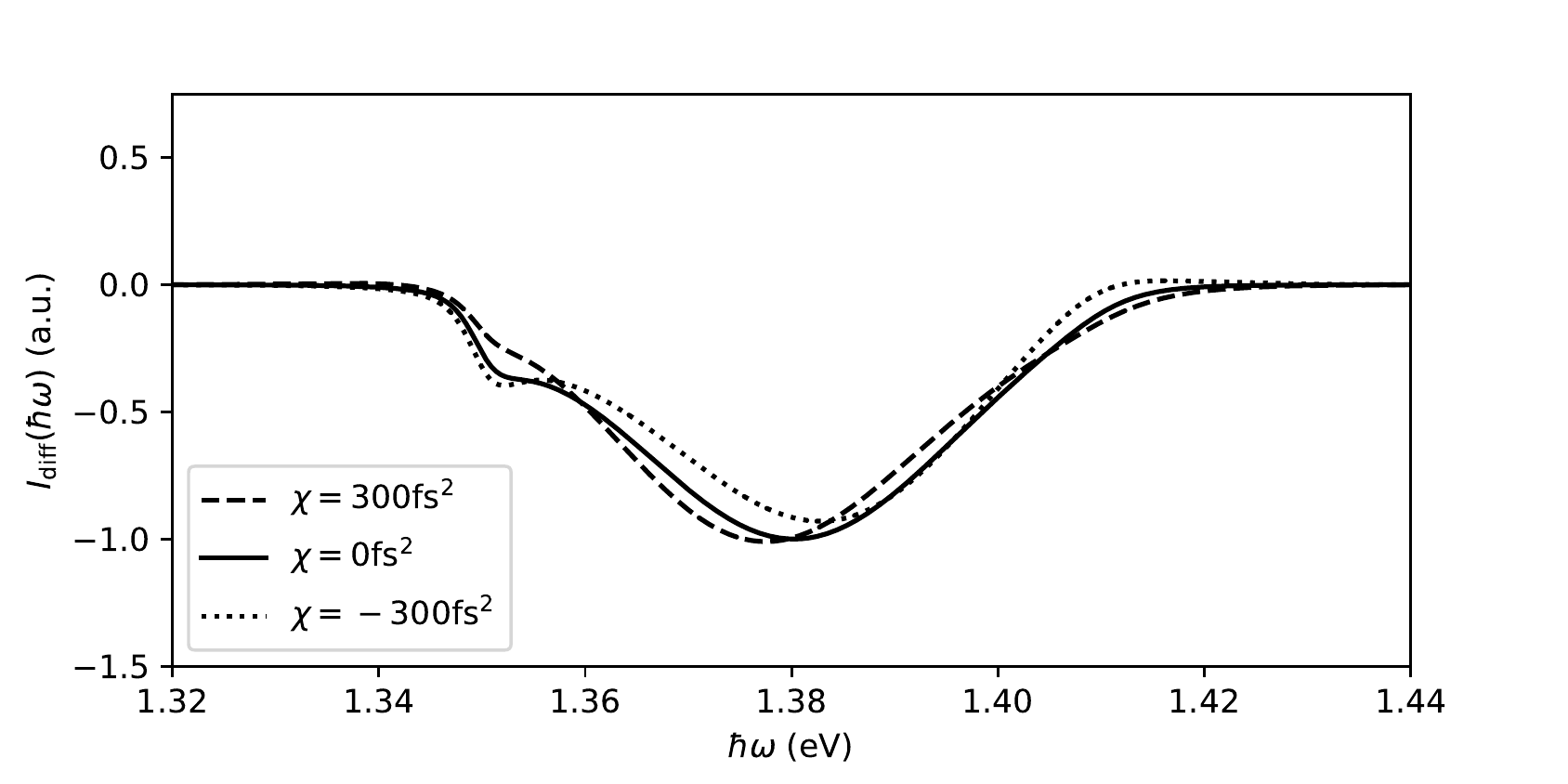}
    \caption{}\label{fig:r3v1_chirps}
  \end{subfigure}
  \begin{subfigure}[b]{0.7\textwidth}
    \includegraphics[width=\textwidth]{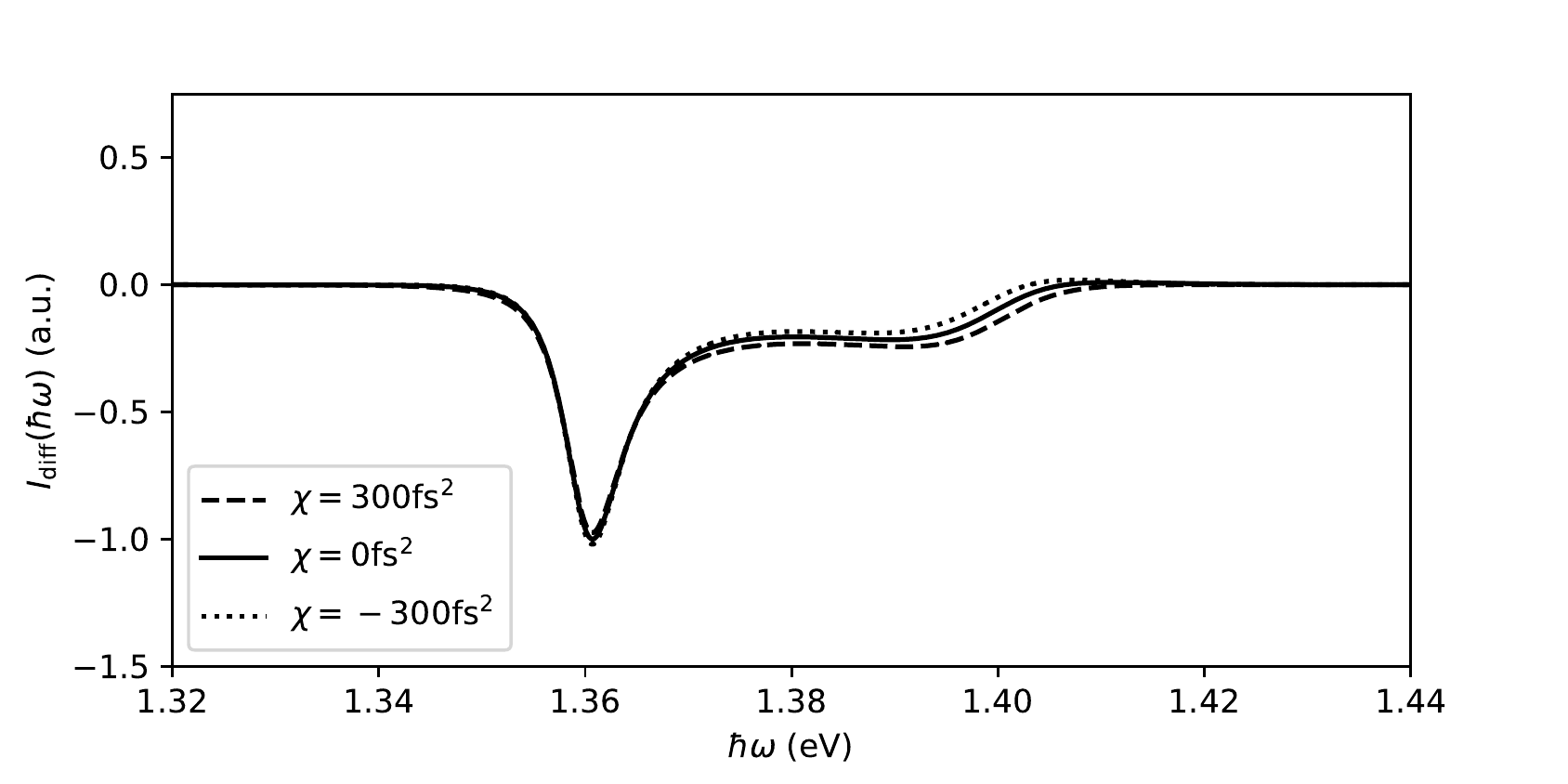}
    \caption{}\label{fig:r2v2_chirps}
  \end{subfigure}
  \caption{
    Change in intensity $I_\text{diff}(\omega)$ for (a) 1 vs. 3
    control, and (b) 2 vs. 2 control at three different values of the
    chirp $\chi$ and at the control intensity $I_0= I_c$.}
\end{figure}

The 1 vs. 3 case, shown in Fig.~\ref{fig:r3v1_spectra} and
\ref{fig:r3v1_chirps}, is complicated by the presence of both a
non-resonant two-photon pathway and the 1 vs. 3 control pathway.  The
far-from-resonance $\ket{v}$ lineshape is responsible for the uniform
absorption of the pulse seen in the weak-field spectrum.  The very
weak $g \rightarrow t$ transition is responsible for the peak at 1.35
eV.\footnote{ The one-photon absorption to $\ket{t}$ is 2500 times
weaker than resonant absorption to $\ket{v}$.}  The increase in
uniform absorption at higher intensity is the result of
non-resonant two-photon absorption through $\ket{v}$.

Given these complications, we note an alternate characterization of the 1 vs. 3 control scheme. Specifically, it is characterized by the chirp
dependence of the one-photon absorption line.  Chirping the pulse, as
is shown in Fig.~\ref{fig:r3v1_chirps}, reveals the control
features. In this case, for a positively chirped pulse, the one-photon
absorption to $\ket t$ interferes destructively with the
pump-pump-dump pathway, even far from saturation.  This is the same
mechanism that is responsible for the dip in quantum yield in
Fig.~\ref{fig:control_qy}.  The same is not seen in 2 vs. 2 control,
shown in Fig.~\ref{fig:r2v2_chirps}.  Since the pulse is too low in
intensity to saturate the $\ket v$ state a change in the chirp does
not significantly affect absorption to $\ket v$.

\subsection{Control of current in ChR2}
Control amplification from repeated pulsed laser interactions is
demonstrated below and the resultant simulation is shown to reproduce the
main qualitative experimental features of control over the peak
current of ChR2.\cite{paul_coherent_2017}  Specifically, a model of
retinal\cite{balzer_mechanism_2003} is used to evaluate the effect of
chirp on the photoisomerization of retinal in ChR2.  The
photoisomerization transition probabilities are then used to compute
transition rates for a classical model of a ChR2 expressing
neuron.\cite{nikolic_photocycles_2009,
  grossman_modeling_2011,evans_pyrho_2016}

The focus of this section is on the 2 vs. 2 coherent control scheme
described above for two reasons: (1) the pump-dump control scenario
with a slow-envelope pulse requires the product energy to be similar
to the ground energy, and (2) the 1 vs. 3 control scheme requires a
delicate balance of the one-photon and three-photon cross-sections.
Specifically, with respect to the former and by analogy with other
opsins, the isomerization product is likely to have a significantly
higher energy than the ground state, since it is this energy storage
that drives changes in protein conformation in e.g.
bacteriorhodopsin.\cite{birge_energy_1983,birge_revised_1991}  If this
is the case also for ChR2, it is unlikely that a direct pump-dump
route exists (or is sufficiently bright) to support a pump-dump
control scheme.  Specifically, with respect to the latter, the 1 vs. 3
control scheme, although possible, is not easily implemented in the
more complex model used here without significant fine-tuning.  Hence,
an analysis of the 1 vs. 3 control scheme awaits further experimental
results.  By contrast, 2 vs. 2 control is robust, requiring only the
presence of a bright near-resonant transitions.  The presence of such
transitions is further supported by the high two-photon absorption
cross-section of
opsins.\cite{birge_twophoton_1990,
  rickgauer_two-photon_2009,palczewska_human_2014}

\subsubsection{Chirp control of retinal isomerization}
Control over the peak current, the quantity of interest here, is the
direct result of coherent control over the photoisomerization of
retinal in ChR2.  At the moderate intensities of interest, the peak
current is directly proportional to the population of ChR2 in the
dark-adapted, highly conductive open state $O_1$, shown in
Fig.~\ref{fig:inco}.  A transition from the initial $C_1$ state to the
open $O_1$ state follows the isomerization of a retinal molecule in
ChR2;\cite{nagel_channelrhodopsin-2_2003,bamann_spectral_2008,
  verhoefen_photocycle_2010,neumann-verhoefen_ultrafast_2013,
  lorenz-fonfria_channelrhodopsin_2014}
thus, the rate of retinal photoisomerization directly determines the
peak current.  Chirp control over photoisomerization is demonstrated in
this section.

\begin{figure}
  \centering
  \begin{subfigure}[b]{\textwidth}
    \includegraphics[width=\textwidth]{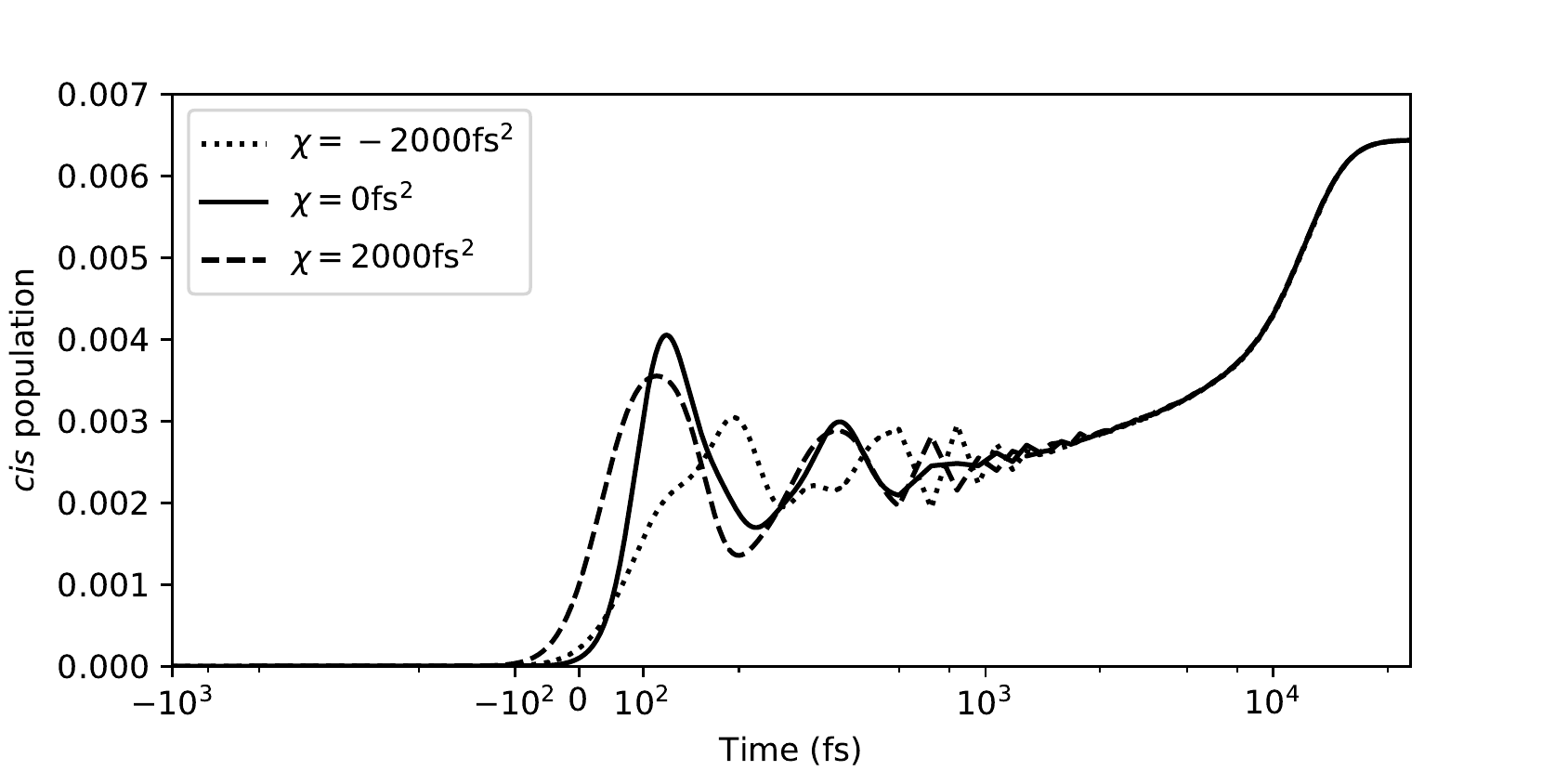}
    \caption{One-photon excitation}\label{fig:one_photon_prop}
  \end{subfigure}
  \begin{subfigure}[b]{\textwidth}
    \includegraphics[width=\textwidth]{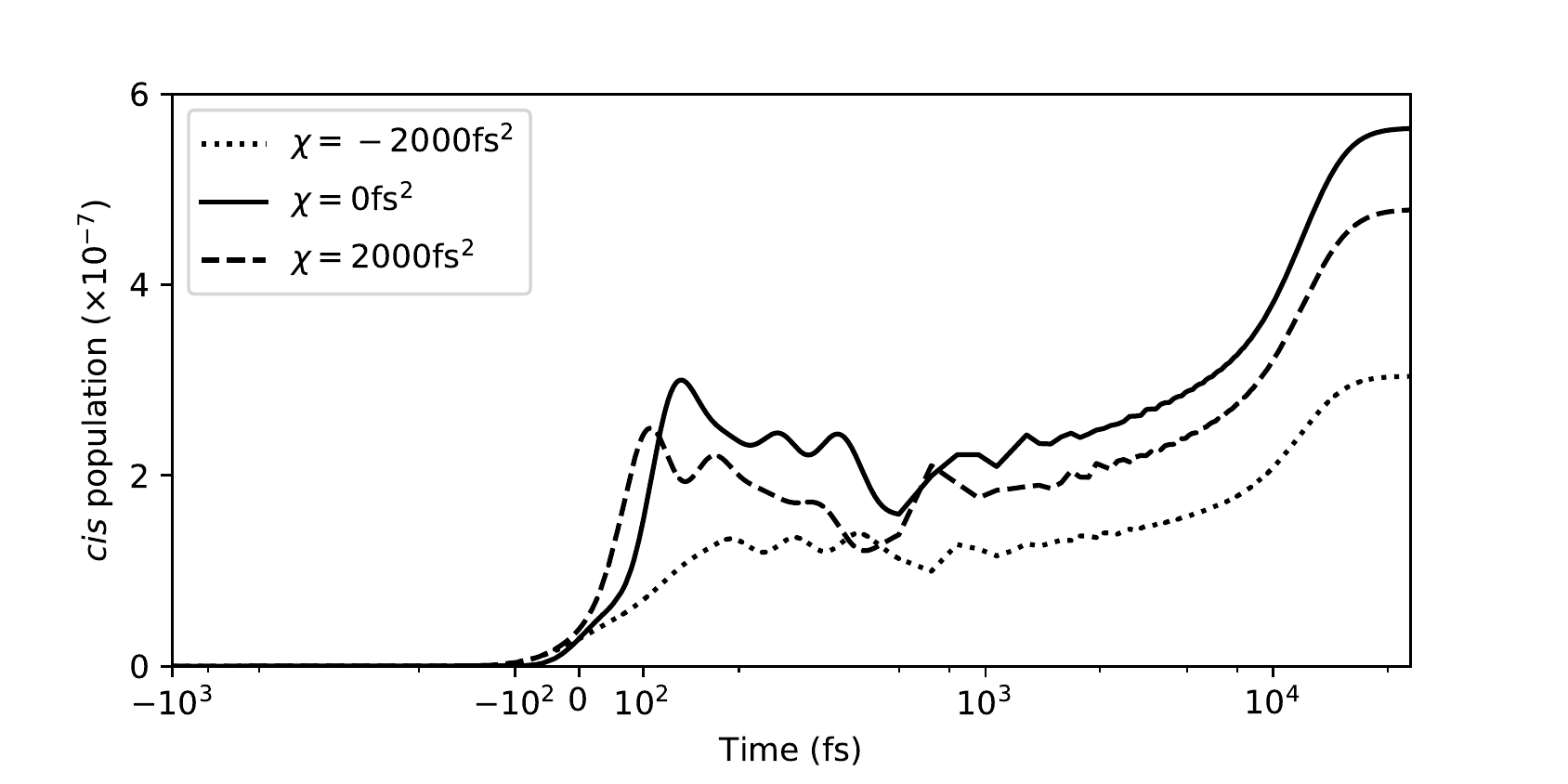}
    \caption{Two-photon excitation}\label{fig:two_photon_prop}
  \end{subfigure}
  \caption{
    Population of the \textit{cis} isomer shortly after
    excitation.  In the (a) one-photon case, the pulse central frequency is
    given by $\hbar \omega = 2.5$ eV, while it is exactly half that in the
    (b) two-photon case.  The FWHM of the unchirped pulse is 120 fs in both
    cases.  Note the difference in ordinate scale in panels (a) and (b).}
\end{figure}

The model of retinal described in Appendix \ref{sec:model-retinal} has
two quasi-steady states, corresponding to \textit{cis} and
\textit{trans} isomers of retinal, with the ground state being of
fully \textit{trans} character.  The result of a one-photon excitation
at 2.5 eV is demonstrated in
Fig.~\ref{fig:one_photon_prop}.  Transient dynamics is seen to be
followed by a ps relaxation to a stable distribution of isomer
populations.  The transient dynamics during and just after the pulse, but not the
steady state population, are seen to be phase dependent.  The
particular form of the bath establishes a high quantum yield of
$\approx 0.75$ (not shown) for the isomerization, consistent with
experiment.

Two-photon chirp control is obtained by modifying this model to
include a near-resonant one-photon transition at half the one-photon
excitation energy.  That is, we added a bright \textit{trans} 
state $\ket{v}$ at 1.22 eV from the ground state (i.e. in the near infrared), thus
creating control of the 2 vs. 2 type.\footnote{
  The $\ket g \rightarrow \ket v$ transition must be resonant to
  obtain chirp control.  Whether its frequency is above or below the
  central frequency of the laser determines whether absorption is
  maximal at negative or positive chirps.  The latter is chosen here to
  match the result of Ref.~\onlinecite{paul_coherent_2017}.}
For simplicity, the state $\ket{v}$ is taken to be vibrationally
identical to $\ket{g}$.  Furthermore, $\ket{v}$ relaxes to $\ket{g}$ in
200 fs; as such, it does not lead to any isomerization.  Such a state
may represent, for example, a highly excited vibrational state from an
anharmonic mode of retinal.  The transition dipole operator is taken to
be of the form,
\begin{align}
  \mu = a_v \mu_{v} + a_e \mu_e \label{eq:mu}
\end{align}
where $\mu_v = \ket{g}\bra{v} + \text{h.c.}$ consists of only the
transition between $\ket{g}$ and $\ket{v}$ and $\mu_e$ includes all
other transitions.

The parameter $a_v$ changes the amplitude of one-photon excitation to
$\ket{v}$ while the parameter $a_e$ sets the transition amplitude of
one-photon excitation to $\ket{e}$ from $\ket{g}$ and from $\ket{v}$.
Hence, changing those parameters modifies the cross-sections for one-
and two-photon absorption and thus the one- and two-photon
isomerization probabilities $P^{(1)}$ and $P^{(2)}$ in
Eq.~\eqref{eq:activation_rate}.  The two isomerization contributions
from Eq.~(\ref{eq:contribs}) obey the following simple relations with
respect to $a_v$ and $a_e$,\footnote{
  One-photon excitations to $\ket{v}$ relax non-productively to
  $\ket{g}$ so that modifying $a_v$ does not change the one-photon
  contribution to isomerization.  In addition, as there are no other
  near-resonant transitions in the model, two-photon absorption
  primarily occurs through $\ket{v}$, such that the two-photon
  absorption probability is proportional to $|a_v|^2$.}
\begin{align}
  P^{(1)} \propto |a_e|^2 \text{ and } P^{(2)} \propto |a_v|^2 |a_e|^2. \label{eq:mu_abs}
\end{align}
Hence, experimental measurements of one- and two-photon absorption
cross-section fully constrain the values of $|a_v|^2$ and $|a_e|^2$,
provided that these measurements are performed at the same near-infrared
frequency.  In the absence of such direct measurements, the near-infrared one-photon
cross-section is estimated from experimental data in
Ref.~\onlinecite{rickgauer_two-photon_2009} (where a similar laser to
Ref.~\onlinecite{paul_coherent_2017} was used) as follows.  The peak
intensity at which two-photon and one-photon excitation rates are the
same is approximately $I_\text{same} \approx 1.7 \times 10^{24} $
photons cm$^{-2}$ s$^{-1}$.  The two-photon cross-section is reported
as $\sigma_2 = 10^{-50}$ cm$^{4}$ s photons$^{-1}$.  Hence, the
one-photon cross-section $\sigma_1$ in the near infrared can be computed from
\begin{align}
  P^{(1)} = I_\text{same} \eta_1 \sigma_1 = P^{(2)} = I^2_\text{same} \eta_2 \sigma_2/2. \label{eq:mu_cross}
\end{align}
where $\eta_1$ and $\eta_2$ are the quantum yield of near-infrared one-photon
and two-photon excitations and $\sigma_1$ is the near-infrared one-photon
cross-section.  As an approximation, the two efficiencies are taken to
be the same ($\eta_1 =\eta_2$), thus fully determining the values of
$|a_e|^2$ and $|a_v|^2$.  We note that at the peak intensity of the
control laser of Ref.~\onlinecite{paul_coherent_2017} ($I_c =5.08
\times 10^{24}$ photons cm$^{-2}$ s$^{-1}$) one-photon contributions
are still significant; that is, contributions from the one-photon
cross-section in the near infrared can not be ignored.

\begin{figure}[h]
    \includegraphics[width=0.5\textwidth]{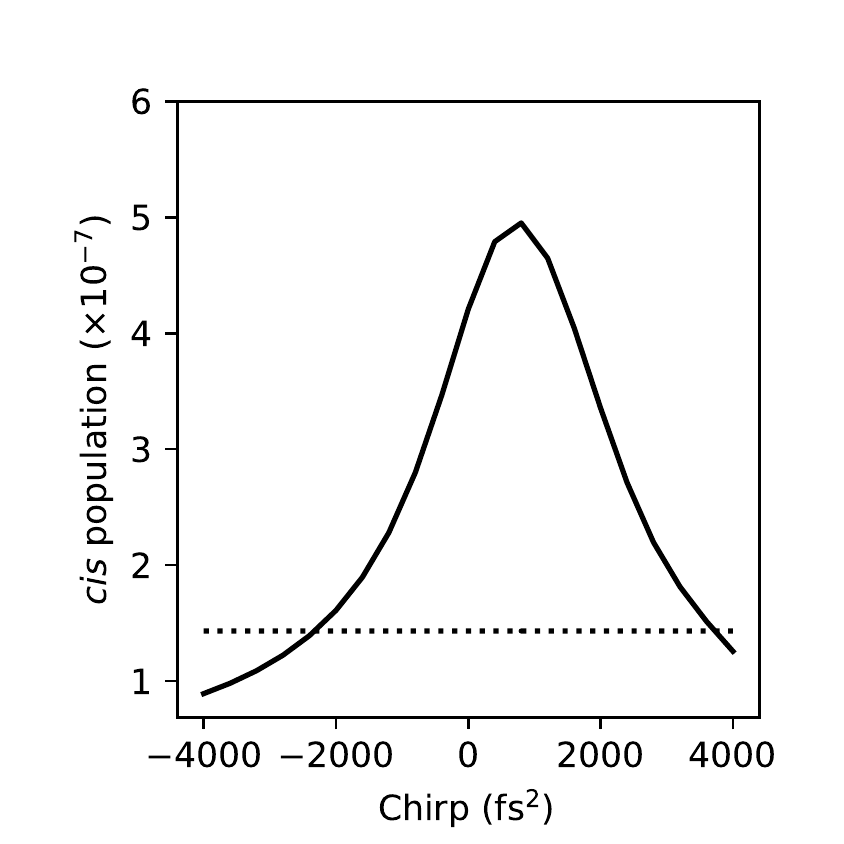}
    \caption{
      (a) Two-photon (solid) and one-photon (dotted)
      contributions $P^{(2)}$ and $P^{(1)}$ to the \textit{cis} population
      after excitation.  The (unchirped) excitation pulse has a FWHM of 120
      fs and a central frequency of $\hbar\omega=1.25$ eV.}
  \label{fig:isomerization}
\end{figure}

Two-photon excitation, as shown in Fig.~\ref{fig:two_photon_prop},
follows similar dynamics to the one-photon excitation of
Fig.~\ref{fig:one_photon_prop}.  This is unsurprising as the $\ket{v}$
state was taken to be vibrationally identical to the ground state.
However, significantly, chirping the pulse now affects \textit{both}
the transient dynamics and the steady state isomer populations.  Since
the state $\ket{v}$ is near resonant, the isomerization is highly
phase controllable (Fig.~\ref{fig:isomerization}), even though less
than one molecules out of a million actually undergoes isomerization.
The computed dependence of the photoisomerization probability on the
chirp of the exciting pulse shown in Fig.~\ref{fig:isomerization} is
similar with respect to the chirp sign and amplitude to that reported
experimentally.\cite{paul_coherent_2017}

\subsubsection{Macroscopic phase control through repeated interactions}
\begin{figure}[h]
  \centering
  \begin{subfigure}[b]{0.4\textwidth}
    \includegraphics[width=\textwidth]{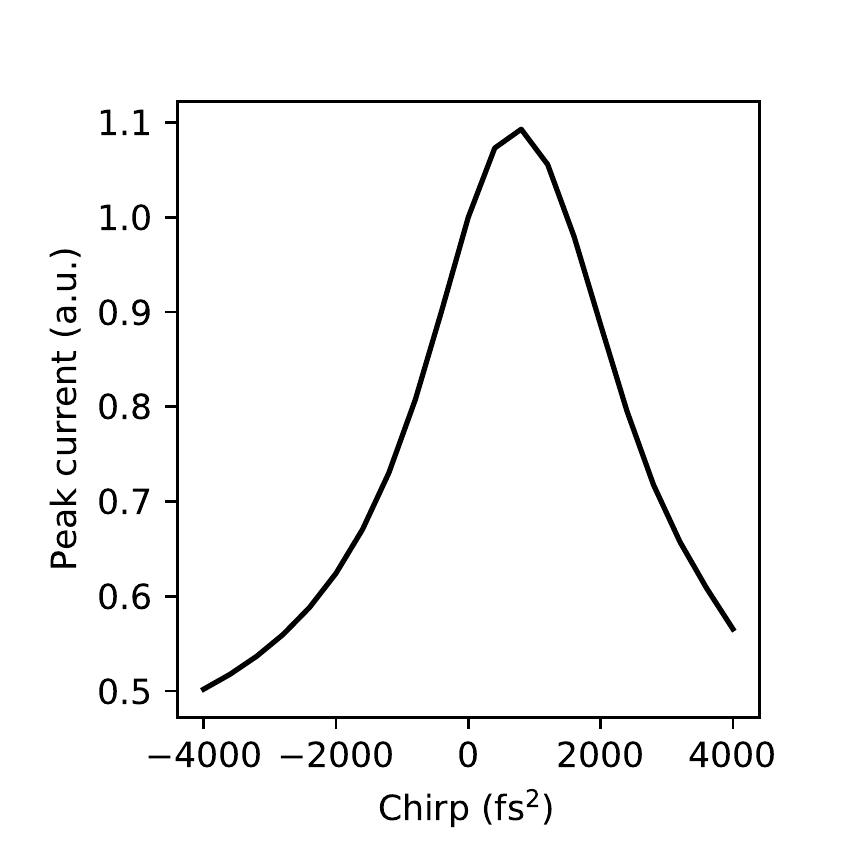}
    \caption{}
    \label{fig:control_current}
  \end{subfigure}
  \begin{subfigure}[b]{0.4\textwidth}
    \includegraphics[width=\textwidth]{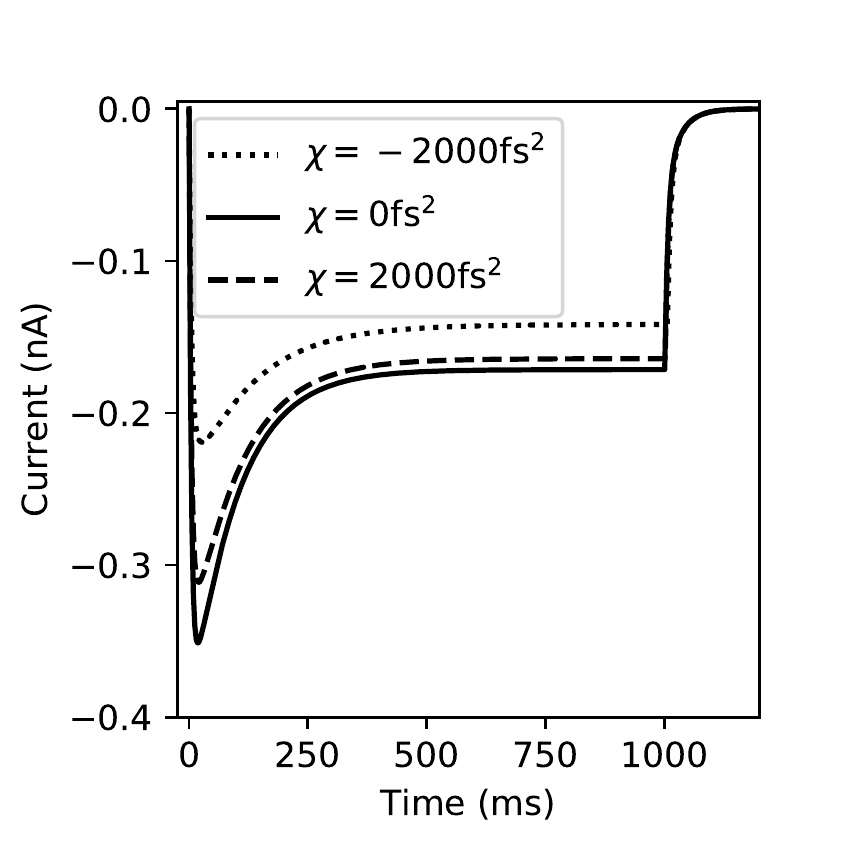}
    \caption{}
    \label{fig:current_td}
  \end{subfigure}
  \caption{
    (a) Peak current as a function of chirp and (b) obtained
    current traces at three different chirp values for the ChR2 model.
    Qualitative agreement is obtained with Figs. 2a and 2c of
    Ref.~\onlinecite{paul_coherent_2017}.}
  \label{fig:control_current_td}
\end{figure}

The single pulse photoisomerization probability obtained above can be
used to compute the $C_1 \rightarrow O_1$ activation rate $K_{a1}$
under illumination with a train of such pulses, as given by
Eq.~\eqref{eq:activation_rate}.  The peak current, at laser powers of
interest, is principally a function of this activation rate, as shown
in Fig.~\ref{fig:inco}.  Thus, the phase control demonstrated above
should be reflected in an equivalent control of the peak current.  The
activation rate $K_{a1}$ was computed using
Eq.~(\ref{eq:activation_rate}) from the isomerization probability due
to a single pulse (Fig.~\ref{fig:isomerization}) and is shown in
Fig.~\ref{fig:control_current}.  The obtained peak current (here
normalized by the value at zero chirp) is seen to depend both on the
magnitude and sign of the applied chirp. {\it The experimental
dependence on the chirp obtained in Figs. 2a and 2c of
Ref.~\onlinecite{paul_coherent_2017} is seen to be well reproduced.}

The non-controllable one-photon excitation (dotted line in
Fig.~\ref{fig:isomerization}) lowers the magnitude of phase control
such that the peak current (Fig.~\ref{fig:control_current}) has a
weaker dependence on the chirp than does the two-photon photoisomerization
probability (Fig.~\ref{fig:isomerization}).  Specifically, at high
magnitude of the chirp, the two-photon contribution decays to zero;
the peak current is then entirely the result of chirp-independent
one-photon photoisomerization.  This was noted by Paul \textit{et al.}
but was attributed to saturation of the light absorption, which is not
the case here, as is demonstrated in the current traces of
Fig.~\ref{fig:current_td}.  That is, the peak current at saturation is
a parameter of the model, here taken to be $I_\text{max} = 2.05$ nA,
far larger than the current that is generated here.  Neither is the
retinal transition saturated, as only a small amount of retinal is
excited by each pulse (Fig.~\ref{fig:two_photon_prop}).  Rather, what
limits the amount of control is the presence of one-photon excitations
(dotted line in Fig.~\ref{fig:isomerization}), which are not
phase-dependent. Significantly, the ratio of one-photon and two-photon
absorption can be modified by changing the repetition rate and
intensity of the laser, hence verifiable experimentally.

\begin{figure}[h]
  \centering
  \begin{subfigure}[b]{0.45\textwidth}
    \includegraphics[width=\textwidth]{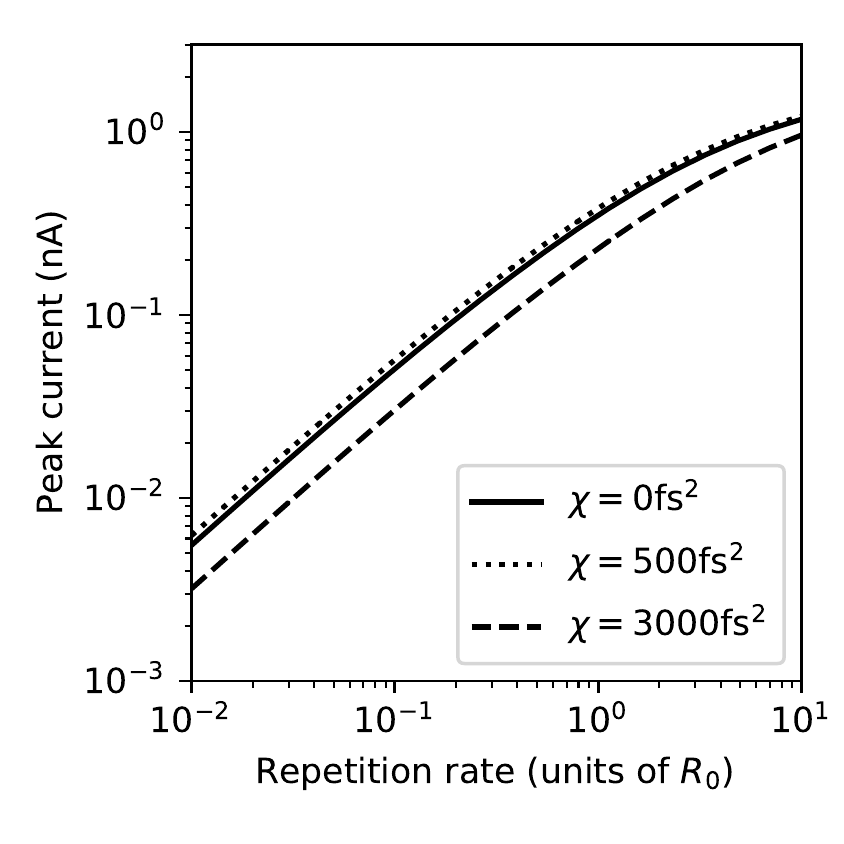}
    \caption{}
      \label{fig:rep_mod}
    \end{subfigure}
  \begin{subfigure}[b]{0.45\textwidth}
    \includegraphics[width=\textwidth]{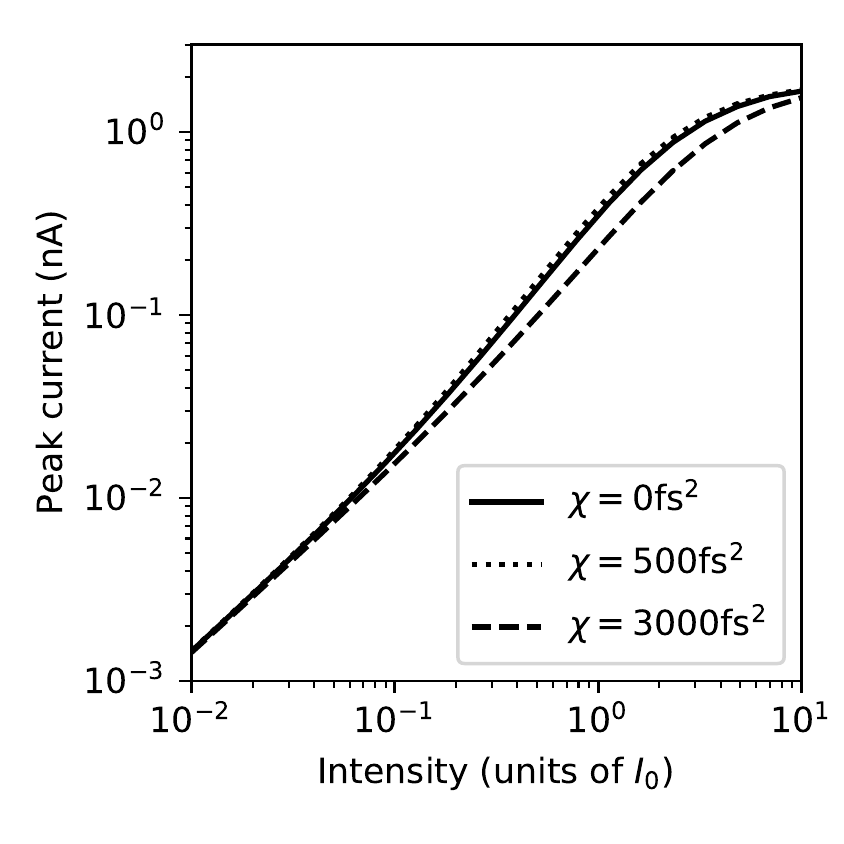}
    \caption{}
      \label{fig:I_mod}
  \end{subfigure}
  \caption{
    Peak current for different values of (a) the repetition rate and
    (b) the peak intensity of the excitation laser, keeping all other
    parameters fixed, at different values of the applied chirp.  $R_0$ and
    $I_0$ denote the repetition rate (80 MHz) and laser peak intensity
    ($5.08 \times 10^{24}$ photons cm$^{-2}$ s$^{-1}$) used in
    Figs.~\ref{fig:isomerization} and \ref{fig:control_current_td}. Note
    the logarithmic scale of both the ordinate and the abscissa.}
  \label{fig:current_mod}
\end{figure}

The amplification of control from repeated interactions is responsible
for the chirp-independent linear relationship between the repetition
rate and the peak current (Fig.~\ref{fig:rep_mod}).  Indeed,
increasing the number of pulses per second produces a corresponding
linear increase in the rate of photoproduct formation, as described by
eq.~(\ref{eq:activation_rate}).  Furthermore, the linear relationship
is phase-independent; a high relative phase control (a significant
change in the photoproduct generated per pulse due to phase control)
can be converted to a high absolute amount of phase control (a
significant change in the actual quantity of photoproduct generated in
the experiment) by increasing the repetition rate.

Changing the repetition rate in this way corresponds to changing the
average laser power while keeping the peak power, and thus the
relative amount of one-photon and multiphoton absorption processes,
fixed.  This is in contrast with the result obtained from changing the
overall laser intensity, as obtained from e.g. applying a neutral
density filter.  Then, the proportion of two-photon and one-photon
absorption processes, and therefore the amount of phase control,
changes nonlinearly, as shown in Fig.~\ref{fig:I_mod}.  Reducing the
intensity leads to a reduction of phase control due to the rapid loss
of two-photon absorption and the lack of phase-sensitivity of the
one-photon terms (Fig.~\ref{fig:one_photon_prop}).  This effect is in
agreement with the experiment of
Ref.~\onlinecite{paul_coherent_2017}.  Specifically, the peak current
was found to be a function $I^\alpha$ of the laser intensity $I$, with
$\alpha$ less than the two-photon expected value of two
(Ref.~\onlinecite{paul_coherent_2017}, Supplemental Fig.~6).  A reduction in intensity was
also shown to dampen the effect of the chirp on the peak current.
Significantly, both results are readily explained by the amplification
mechanism and the presence of a significant, phase-independent
one-photon contribution. In addition, this mechanism is valid in the
regime where neither the absorption of retinal nor the dynamics of
ChR2 are close to saturation, which is consistent with the scaling
study of Ref.~\onlinecite{rickgauer_two-photon_2009}, and reproduces,
as described above, the nonzero peak current detected at high
magnitude of the chirp.

Thus, repeated weak interactions were shown here to generate large phase
controllable currents.  Significantly, control amplification can be
experimentally distinguished from other processes by measuring the
magnitude of phase dependence at different repetition rates and laser
intensities.  Furthermore, by increasing the laser intensity while
simultaneously decreasing the laser repetition rate, control can be
made stronger without saturating either the initial absorption process
or the generation of current.  The same protocol can be applied to
other multiphoton photochemical processes induced ultrafast pulsed
lasers.

\section{Conclusion}
A theory for two-photon phase control of a macroscopic phenomena, such
as the experimental control by light of electrical current in live
neurons, has been described.  Three microscopic mechanisms based on
the properties of the exciting field have been proposed with
measurable spectroscopic signatures.  Under certain assumptions,
microscopic control provided by individual pulses in a pulse train
accumulates, leading to a large phase effect, which may persist over
macroscopic timescales.

A model for the experiment of Paul \textit{et al.} on the coherent
control of electrical currents emanated from living brain tissue
expressing ChR2, was proposed.\cite{paul_coherent_2017}  Specifically,
a quantum mechanical model of retinal\cite{
  hahn_quantum-mechanical_2000, hahn_ultrafast_2002,
  balzer_mechanism_2003}
with experimentally derived one- and two-photon absorption
cross-sections\cite{rickgauer_two-photon_2009} was used to compute
quantum mechanical isomerization probabilities.  These were applied to
derive activation rates for ChR2 under repeated pulsed excitation.  The
ms current dynamics of ChR2-expressing neurons were computed using a
set of rate equations.\cite{nikolic_photocycles_2009,
  grossman_modeling_2011,evans_pyrho_2016}
This multi-timescale model was shown to reproduce the dependence of
the peak current on the chirp of a pulsed laser excitation observed
experimentally.  The microscopic mechanism of control is consistent
with interference between two-photon excitation routes (i.e. 2 vs. 2
photon control) arising from a near resonant transition.

The microscopic mechanism and subsequent control amplification
processes proposed here are experimentally verifiable.  The control
scheme requires the presence of measurable bright near-resonant
transitions and has a specific spectral signature
(Fig.~\ref{fig:r2_spectra}).  Control amplification through repeated
interactions is testable by changing excitation parameters, namely the
repetition rate and the intensity of the laser; the effect of chirp on
the peak current depends only on the latter
(Fig.~\ref{fig:current_mod}).  That is, the ratio of peak current obtained at different $\chi$ is relatively constant as a function of the repetition rate which is not the case for changes in laser intensity.

Other features of the two-photon excitation of
ChR2\cite{rickgauer_two-photon_2009,paul_coherent_2017} are readily
accounted for by the present model: the peak current is a function of
the intensity $I^\alpha$, with $\alpha$ between one (linear) and two
(quadratic), the effect of the chirp on the peak current is dampened
when the intensity is reduced, and the peak current does not fully
vanish at high values of the chirp.  These features also appear in this
model study and are explained as the result of nonzero
phase-independent one-photon excitations.\cite{
  brumer_one_1989, spanner_communication_2010,
  liebel_lack_2017,lavigne_ultrafast_2019}
The one-photon contribution can be directly measured by increasing the
laser intensity while reducing the repetition rate, i.e. by reducing
the laser peak power while keeping fixed the laser average power.

Future work should include more realistic treatment of retinal in ChR2
and of neuron current dynamics.  The simulation presented here is
primarily a qualitative proof of concept for the described control
mechanisms.  A quantitative treatment necessitates a more advanced
model of retinal in ChR2, informed, e.g., by near-infrared spectroscopy
data.\cite{neumann-verhoefen_ultrafast_2013}  In particular, two-photon
absorption was included here somewhat artificially; more complete data
on one-photon and two-photon near-infrared absorption in ChR2,
including which vibrational modes and which electronic states are
active, would greatly improve modeling.  The role of the retinal
environment is of particular theoretical interest.  Significantly, the
Redfield treatment used here imposes too broad Lorentzian linewidths,
thus exaggerating absorption far from resonances.  Large system closed
dynamics\cite{
  vendrell_multilayer_2011,christopher_efficient_2006,
  arango_communication_2013,lavigne_qp_2014}
and non-Markovian
methods\cite{shao_iterative_2002,
  pachon_mechanisms_2013, ma_forster_2015,
  ma_forster_2015-1,moix_forster_2015}
can be used to obtain more realistic, e.g. Gaussian, lineshapes.
Finally, a treatment of the neuron biophysics which includes
activation dynamics\cite{grossman_modeling_2011} is required to obtain
the other phase effect reported in
Ref.~\onlinecite{paul_coherent_2017}, i.e., that the chirp of the
exciting laser changes the spiking pattern of neurons.

The use of more complex pulse shapes to selectively activate specific
opsins is particularly interesting for optogenetics applications.  For
example, different neurons can be made to express different opsins,
but selective excitation of opsins is currently limited by a high
degree of spectral overlap between different
species.\cite{schneider_biophysics_2015}  Phase control provides an
entirely new dimension beyond simple spectral control, which could
generate significantly better discrimination between types of opsins,
as was previously demonstrated for two-photon induced
fluorescence.\cite{pastirk_selective_2003,schelhas_advantages_2006,
  ogilvie_use_2006,tkaczyk_control_2009,
  isobe_multifarious_2009,brenner_two-photon_2013}
Another avenue for selective excitation is through spatiotemporal
control, where both the spatial and temporal phase of light are
modulated.\cite{katz_focusing_2011,mounaix_spatiotemporal_2016}  Joint
spatiotemporal focusing and two-photon phase control may be of use in
high resolution targeting of individual neurons and
dendrites.\cite{papagiakoumou_scanless_2010}

The present analysis can also be applied to other photochemical
processes where fast two-photon excitation generate macroscopically
long-lived photoproducts, e.g. the photoisomerization of
azobenzenes,\cite{izquierdo-serra_two-photon_2014,
  carroll_two-photon_2015,wei_two-photon_2019}
the excitation of phosphorescent
molecules\cite{sakadzic_two-photon_2010} and the photodissociation of
molecular species.\cite{scaiano_photochemistry_1988,
  pastirk_quantum_1998,potapov_two-photon_2001}
The repeated application of weak pulses can rapidly create a large
phase-controllable amount of photoproducts in the manner described
here, provided certain conditions are met.  Denoting the time between
pulses by $\tau$, a similar treatment to this one is possible when (1)
the microscopic timescales of coherence and product formation are much
shorter than $\tau$, (2) the macroscopic timescales of interest are
much longer than $\tau$, and (3) only a fraction of the system is
excited by each individual pulse.  For a MHz laser repetition rate,
complete formation of products and decoherence must occur within 1
$\mu$s of excitation, while other macroscopic processes, e.g. decay or
removal of photoproducts, must have characteristic timescales much
longer than a $\mu$s.  This type of separation of timescales is common
in photochemical and photobiological processes.

\textbf{Acknowledgments:} This work was supported by the
U.S. Air Force Office of Scientific Research under Contract No.
FA9550-17-1-0310, and by the
Natural Sciences and Engineering Research Council of Canada.


\appendix

\section{Current dynamics of ChR2-expressing neurons}\label{sec:model-current}
The macroscopic current produced by neurons expressing
channelrhodopsin-2 is evaluated using a classical rate model, with
empirically determined rates taken from
Ref.~\onlinecite{nikolic_photocycles_2009} and
Ref.~\onlinecite{evans_pyrho_2016}.  The four-state conductance model
(Fig.~\ref{fig:rate_model}) is simulated using the following rate
equations,
\begin{align}
  \ddt{C_1(t)} &= -K_{a1}C_1(t) + K_r C_2(t) + K_{d1} O_1(t)\\
  \ddt{C_2(t)} &= -(K_{a2} + K_r) C_2(t) + K_{d2} O_2(t)\\
  \ddt{O_1(t)} &= K_{a1} C_1(t) - (K_{d1} + e_{12}) O_1(t) + e_{21} O_2(t)\\
  \ddt{O_2(t)} &= K_{a2} C_2(t) - (K_{d2} + e_{21}) O_2(t) + e_{12} O_2(t).
\end{align}
The rates for each process are given in Table~\ref{tab:currparams}.  The
neuronal current is obtained from
\begin{align}
  I(t) = I_\text{max}(O_1(t) + s O_2(t)),
\end{align}
where $s$ is the ratio of the conductivity of the light- and
dark-adapted open states.

The activation rates $K_{ai}$ of the model for an incoherent blue
light source were obtained by a fit to the empirical Hill
equation,\cite{evans_pyrho_2016}
\begin{align}
 K_{ai} = k_i \frac{\Phi^p}{\Phi^p + \Phi_m^p} \label{eq:hill}
\end{align}
where $i = 1,2$, $\Phi$ is the photon flux in m$^{-2}$ s$^{1}$, $p$ is
a parameter near unity and $k_i$ is has units of time.  This nonlinear
equation is used to reproduce nonlinear saturation effects at high
intensity.  This nonlinear form is not based on the physics of
light-matter interactions.  As this treatment focuses on the weak
field regime away from saturation, the following physically-motivated
linear form is used instead,
\begin{align}
  K_{a1} = \eta_1 \sigma_\text{vis} \Phi \text{ and } K_{a2} = \eta_2 \sigma_\text{vis} \Phi.
\end{align}
where $\eta_1$ and $\eta_2$ are the quantum yields of dark- and
light-adapted ChR2, $\Phi$ is the photon flux and $\sigma $ is the
cross-section of retinal, assumed here to be the same for both
states. At the intensity of interest, Eq.~(\ref{eq:hill}) is nearly
linear for both dark- and light-adapted states; a linear fit of this
equation provides the quantum yields $\eta_1$ and $\eta_2$ given in
Table~\ref{tab:currparams} and the visible-light cross-section
$\sigma_\text{vis} = 8\times 10^{-8} \mu$m$^{-2}$.  For two-photon
absorption, the dark-adapted activation rate $K_{a1}$ is computed from
the retinal model using Eq.~(\ref{eq:activation_rate}).  The
light-adapted activation rate, which has a negligible impact on the
peak current, is estimated here from the zero chirp dark-adapted
activation rate $K^{(0)}_{a1}$,
\begin{align}
  K_{a2} = \frac{\eta_2}{\eta_1} K^{(0)}_{a1}.
\end{align}

\begin{table}
\begin{ruledtabular}
  \begin{tabular}{llllll}
              & $K_{d1}$ & $K_{d2}$ & $e_{12}$ & $e_{21}$  & $K_r$ \\
    light off$^\dagger$ & 130 s$^{-1}$ &  25 s$^{-1}$ &  22  s$^{-1}$ & 11 s$^{-1}$ & 0.4 s$^{-1}$ \\
    light on$^*$  & "       &  "      &  53 s$^{-1}$ & 23 s$^{-1}$   & "  \\
    \hline
              & $\eta_1$  &  $\eta_2$ & s       & $I_\text{max}$\\
    Both$^\dagger$     & 0.50    &  0.10    & 0.05    & -1.85 nA\\
\end{tabular}
\end{ruledtabular}
  {\footnotesize
    $\dagger$: Ref.~\onlinecite{nikolic_photocycles_2009}, inset of Fig. 8.\\
    $*$: ibid., p. 408.
  }
  \caption{
    Parameters of the macroscopic model of ChR2 used to compute
    currents produced by neurons after two-photon excitation.  All
    parameters were taken from
    Ref.~\onlinecite{nikolic_photocycles_2009}.}
  \label{tab:currparams}
\end{table}

\section{Perturbative description of weak-field two-photon absorption and numerical evaluation}\label{sec:pertub}
In this section, a derivation for Eq.~(\ref{eq:rho_pt2}) and
(\ref{eq:rho_pt4}) is provided.  The starting point of this analysis is
the Liouville equation given by Eq.~(\ref{eq:lvneq}).  The following
time-dependent Green's function is defined,
\begin{align}
  \super G_0(t-t_0) = \Theta(t-t_0) e^{\super L_0 (t-t_0)}~,
\end{align}
where $\Theta(t)$ is the Heaviside step function.  A perturbative
expansion to fourth order in the field is necessary to expose all
processes quadratic in the field intensity.  The terms of the
perturbative series are given by
\begin{align}
  \rho_0(t) &= \super G_0(t-t_0)\rho(t_0)\\
  \rho_1(t) &= \left(\frac{i}{\hbar}\right)^1  \int_{t_0}^{\infty} \mathrm d t_1 \field(t_1)  \super G_0(t-t_1) \super V \rho_0(t_1)\\
  \rho_2(t) &= \left(\frac{i}{\hbar}\right)^2 \iint_{t_0}^{\infty}\mathrm d t_2\mathrm d t_1 \field(t_2) \field(t_1) \super G_0(t-t_2) \super V \super G_0(t_2-t_1) \super V \rho_0(t_1)\nonumber\\
  \cdots \nonumber
\end{align}
Here, the state at $t_0$ before the field is on is taken to be a
steady state, such that $\super G_0(t-t_0) \rho(t_0) = \rho_0$.  The
lower bound $t_0$ of the integrals is extended to $-\infty$.  The field
is expanded into its Fourier components to obtain,
\begin{align}
  \rho_1(t) &= \left(\frac{i}{\hbar}\right)^1  \int_{-\infty}^{\infty} \mathrm d \omega_1 \field(\omega_1)  \int_{-\infty}^{\infty}\mathrm d t_1 e^{i\omega_1 t_1}\super G_0(t-t_1) \super V \rho_0\\
  \rho_2(t) &= \left(\frac{i}{\hbar}\right)^2  \iint_{-\infty}^{\infty}\mathrm d \omega_2\mathrm d \omega_1 \field(\omega_2) \field(\omega_1) \iint_{-\infty}^{\infty}\mathrm d t_2\mathrm d t_1 e^{i\omega_2 t_2 + i\omega_1 t_1}\super G_0(t-t_2) \super V \super G_0(t_2-t_1) \super V \rho_0\nonumber\\
  \cdots \nonumber
\end{align}
The Fourier integrals thus exposed can be used to transform the
Green's functions.  Certain properties of the field, applicable to
ultrashort pulses, are necessary for this transform to be
well-behaved.

Consider the following integral,
\begin{align}
  F(t_2) &= \int_{-\infty}^{\infty} \mathrm d \omega_1 \field(\omega_1)  \int_{-\infty}^{\infty}\mathrm d t_1 e^{i\omega_1 t_1}\super G_0(t_2-t_1)\label{eq:ft2}\\
           &=e^{i\omega_1 t_2}\int_{-\infty}^{\infty} \mathrm d \omega_1 \field(\omega_1)  \int_{0}^{\infty}\mathrm d \tau e^{-(i\omega_1 - \super L_0) \tau }
\end{align}
where $\tau = t_2 - t_1$.  The second integral converges provided that
the spectrum of $(i\omega_1 - \super L_0)$ is entirely positive.  If
that is the case, the following is obtained,
\begin{align}
         F(t_2)  &= e^{i\omega_1 t_2 }\int_{-\infty}^{\infty} \mathrm d \omega_1 \field(\omega_1)\frac{1}{i\omega_1  - \super L_0 }\label{eq:greens_w}
\end{align}
The eigenspectrum of $\super L_0$ is entirely imaginary for a closed
system, with eigenvalues $i\omega_{nm}$ for every coherence obtained
from a pair of eigenenergies $n,m$ and zero eigenvalues for every
population of an eigenenergy.  Redfield or Lindblad equations contain
decaying contribution (due to decoherence and dissipation); the
corresponding Liouvillian thus has eigenvalues with negative real
parts.  Hence, all the poles of $[i\omega_1 - \super L_0]^{-1}$ in
Eq.~\ref{eq:greens_w} are contained in the upper half plane and on the
real axis.  The latter poles are easily avoided if the field is
time-limited.

\begin{figure}[h]
  \includegraphics{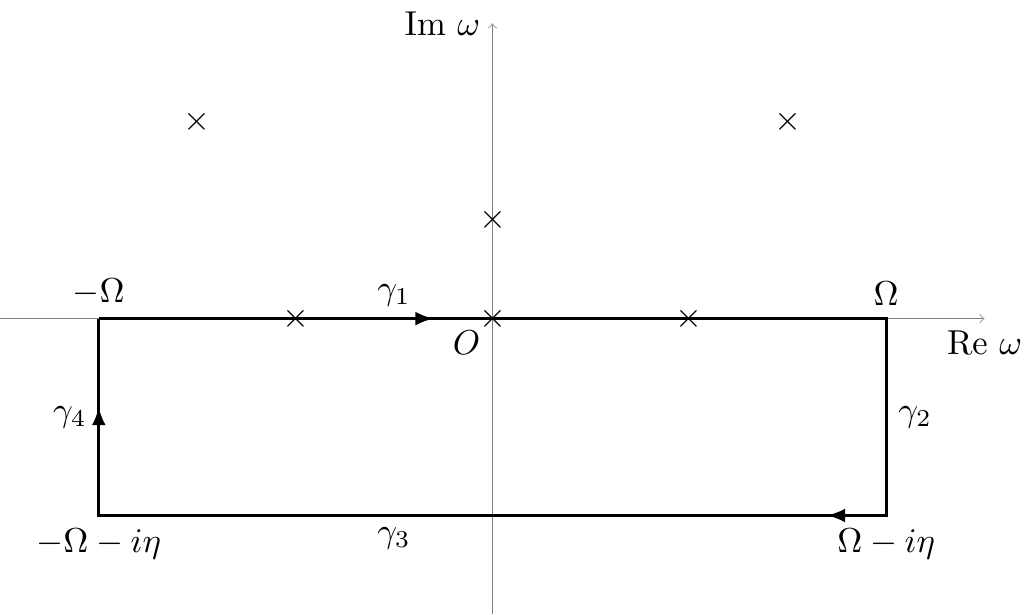}
  \caption{
    Contour integral for Eq.~(\ref{eq:ft2}). $\super
    G_0(\omega_1)$ has poles, represented by crosses, in the upper
    half-plane for an open system or along the $\text{Re}\,\omega_1$ axis
    for a closed system.  The value of $\eta >0$ is a matter a numerical
    convergence.\cite{lavigne_efficient_2019}}
  \label{fig:contour}
\end{figure}

Consider a field $\field(t)$ compactly supported in the time-domain
(time-limited), i.e.
\begin{align}
  \field(t) = 0 \quad \text{for } |t-t_1| > T/2
\end{align}
where $t_1$ defines the center of an interval of size $T$ over which
the field is nonzero.  Then, by the Paley-Wiener
theorem,\cite{rudin_real_1987} $\field(\omega)$ is an entire function
over complex $\omega$ and is square-integratable over horizontal
lines, 
\begin{align}
  \int_{-\infty}^\infty \mathrm d \omega |\field(\omega-i\eta)|^2 < C.
\end{align}
The integral of Eq.~(\ref{eq:ft2}) is given by $\gamma_1$ in
Fig.~\ref{fig:contour}.  By the Cauchy integral theorem,
\begin{align}
  \gamma_1 + \gamma_2 + \gamma_3 + \gamma_4 = 0.
\end{align}
As the field is square-integratable, $\gamma_2$ and $\gamma_4$ are zero, such that $\gamma_1 = -\gamma_3$.  Thus, the integration of
$\omega_1$ from $-\infty$ to $\infty$ can instead be performed over
the line from $-\infty -i\eta$ to $\infty -i\eta$, where $\eta$ is a
positive real number, thereby avoiding all poles of $\super
G_0(\omega_1)$, provided that the field is time-limited.  The following
is then obtained,
\begin{align}
  \rho_1(t) &= \left(\frac{i}{2\pi\hbar}\right)^1  \int_{-\infty}^{\infty} \mathrm d \omega_1 \field(\omega_1-i\eta)  \int_{-\infty}^{\infty}\mathrm d t_1 e^{i\omega_1 t_1 + \eta t_1}\super G_0(t-t_1) \super V \rho_0\\
  \rho_2(t) &= \left(\frac{i}{2\pi\hbar}\right)^2  \iint_{-\infty}^{\infty}\mathrm d \omega_2\mathrm d \omega_1 \field(\omega_2) \field(\omega_1-i\eta) \iint_{-\infty}^{\infty}\mathrm d t_2\mathrm d t_1 e^{i\omega_2 t_2 + i\omega_1 t_1 + \eta t_1 }\super G_0(t-t_2) \super V \super G_0(t_2-t_1) \super V \rho_0\nonumber\\
  \cdots \nonumber
\end{align}
These integrals can be evaluated by applying Eq.~(\ref{eq:ft2}) to
each time integral in sequence.  The first order term (where $t_2=t$)
is given by,
\begin{align}
  \rho_1(t) &= \left(\frac{i}{2\pi\hbar}\right)^1  \int_{-\infty}^{\infty} \mathrm d \omega_1 \field(\omega_1-i\eta)   e^{i\omega_1 t + \eta t}\super G_0(\omega_1-i\eta) \super V \rho_0.\label{eq:rho_pt1_appendix}
\end{align}
Some algebra yields the following higher order terms,
\begin{align}
  \rho_2(t) &= \left(\frac{i}{2\pi\hbar}\right)^2  \iint_{-\infty}^{\infty}\mathrm d \omega_2\mathrm d \omega_1 e^{i(\omega_2 + \omega_1) t + \eta t}\field(\omega_2) \field(\omega_1-i\eta) \\
  &\quad \times \super G_0(\omega_2 + \omega_1 -i\eta) \super V \super G_0(\omega_1 - i\eta) \super V \rho_0\nonumber\\
  \rho_3(t) &= \left(\frac{i}{2\pi\hbar}\right)^3  \iiint_{-\infty}^{\infty}\mathrm d \omega_3 \mathrm d \omega_2\mathrm d \omega_1 e^{i(\omega_3 + \omega_2 + \omega_1) t + \eta t}\field(\omega_3) \field(\omega_2) \field(\omega_1-i\eta) \\
  &\quad \times \super G_0(\omega_3 + \omega_2 + \omega_1 -i\eta) \super V \super G_0(\omega_2 + \omega_1 -i\eta) \super V \super G_0(\omega_1 - i\eta) \super V \rho_0\nonumber\\
  \rho_4(t) &= \left(\frac{i}{2\pi\hbar}\right)^4  \iiiint_{-\infty}^{\infty}\mathrm d \omega_4 \mathrm d \omega_3 \mathrm d \omega_2\mathrm d \omega_1 e^{i(\omega_4 + \omega_3 + \omega_2 + \omega_1) t + \eta t}\field(\omega_4) \field(\omega_3) \field(\omega_2) \field(\omega_1-i\eta)\nonumber\\
          &\quad \times\super G_0(\omega_4 + \omega_3 + \omega_2 + \omega_1 -i\eta) \super V \super G_0(\omega_3 + \omega_2 + \omega_1 -i\eta)\label{eq:rho_pt4_appendix}\\
            &\quad \times \super V \super G_0(\omega_2 + \omega_1 -i\eta) \super V \super G_0(\omega_1 - i\eta) \super V \rho_0\nonumber\\
  &\cdots \nonumber
\end{align}
Every term of the perturbative expansion converges when the field is
time-limited, as is the case here.

The effect that the system has on the outgoing electric field is
computed below using the microscopic equations, under the assumption
that the sample is highly dilute.\cite{mukamel_principles_1995}
Consider the change in the energy of the controlled system due to
radiation,
\begin{align}
  \Delta E_s(t) = - \frac{\mathrm d \field(t)}{\mathrm d t} \Tr \mu \rho(t)
\end{align}
Using the Fourier representation of the field, the following is
obtained,
\begin{align}
  \Delta E_s(t) &= \frac{1}{2\pi i} \int_{-\infty}^{\infty}\mathrm d \omega (-\omega) \field^*(\omega)  \Tr \mu \rho(t) e^{-i\omega t}
\end{align}
A gain of energy by the system must be balanced by a corresponding
loss in the electric field.  The overall change in the energy of
radiation after interacting with the system is then given by,
\begin{align}
  \Delta E_r = -\int_{-\infty}^{\infty} \mathrm d t \Delta E_s(t) 
\end{align}
The $n$-th order contribution in the electric field can then be
obtained from the $n-1$ density matrix,
\begin{align}
   \Delta E_r^{(n)} = \frac{1}{2\pi i} \int_{-\infty}^{\infty}\mathrm d \omega \,\omega \field^*(\omega)  \Tr \mu \rho^{(n-1)}(\omega)
\end{align}
The change in the overall energy of the field after encountering a
single molecule is expressed as a change in the number of photons in
mode $\omega$ times the energy of photons in that mode,
\begin{align}
   \Delta E_r^{(n)} =  \int_{-\infty}^{\infty}\mathrm d \omega N^{(n)}_\Delta (\omega) \hbar \omega  
\end{align}
with the following spectrally-resolved changed in the photon flux due
to interactions of order $n$,
\begin{align}
  N_\Delta^{(n)}(\omega) = \frac{1}{2\pi \hbar i} \field^*(\omega)  \Tr \mu \rho^{(n-1)}(\omega).
\end{align}
The change in the spectral photon flux $N_\Delta(\omega)$ is the sum
of the second and fourth order contributions $N_\Delta^{(2)}(\omega) +
N_\Delta^{(4)}(\omega)$.

The weak-field perturbative expansion
[Eqs. (\ref{eq:rho_pt1_appendix}) to (\ref{eq:rho_pt4_appendix}) has a
few important features, which warrants its use for numerical work as
well as for analytical calculations, even though it is more
complicated than a direct numerical computation.  Indeed, perturbation
theory can be avoided numerically by simply evolving the semiclassical
Liouvillian given by Eq.~(\ref{eq:lvneq}) directly.  However, the
direct method is problematic for three reasons: (a) at higher
excitation intensities, the density matrix may contain undesired
higher order contributions; (b) at lower excitation intensities
numerical errors can drastically affect the very small excited state
populations, requiring fine numerical tolerances, and (c) measuring
the scaling of control with intensity requires repeated computations,
which can rapidly (due to the previous two issues) become
computationally expensive and numerically difficult.  These issues are
avoided by a numerical computation of the perturbative
series.  Furthermore, the absorption spectrum and other nonlinear
spectroscopic properties are more easily calculated perturbatively,
especially when multiple spatial or polarization components of the
field are included.\cite{mukamel_principles_1995}

The numerical evaluation of Eqs.~(\ref{eq:rho_pt1_appendix}) to
(\ref{eq:rho_pt4_appendix}) is difficult due to the
high-dimensionality of the frequency integrations.  As seen above, the
perturbative expansion of $\rho(t)$ for weak-field two-photon
processes requires the evaluation of a four-dimensional frequency
integrals.  Additionally, the
Green's function above may have poles on the real frequency axis,
further frustrating numerical efforts.  For these reasons, we have
developed the algorithm described in
Ref.~\onlinecite{lavigne_efficient_2019}, which avoids all of these
issues.  This numerical method was successfully used to obtain the
results described in this paper.

\section{Minimal models for two-photon phase control} \label{sec:minimal-models}
Three control models are used in Sec.~\ref{sec:mech-two-phot}.  The
models consists of five levels (Fig.~\ref{fig:model}).  Dissipation
and decoherence are included to obtain sub-unity quantum yields and
continuous lineshapes.  The following Lindblad relaxation tensor is
defined,
\begin{align}
  \super R_L\rho = \sum_k \gamma_k\left( L_k \rho L_k^\dagger - \frac{1}{2}\left\{L_k^\dagger L_k,
  \rho\right\}\right)
\end{align}
The material Liouvillian of Eq.~(\ref{eq:lvneq}) is then given by
\begin{align}
  \super L_0\rho = \frac{1}{i\hbar}\commute{H_0}{\rho}  + \super R_L \rho.
\end{align}
The decoherence terms are given by,
\begin{align}
  L_k = \ket{k}\bra{k}
\end{align}
where $k$ runs over all indices of the field and the corresponding
rate is given by $\gamma_k = 2/\tau_\text{deco}$.  The electronic
excited state has a short electronic decoherence time
$\tau_\text{el,deco}$ while the remaining states are assigned a slower
vibrational decoherence time $\tau_\text{vib,deco}$.  The dissipation
operator for the $i\rightarrow j$ relaxation has the form,
\begin{align}
  L_{i\rightarrow j} = \ket{j}\bra{i}.
\end{align}
The dissipation rate for the $\ket{v}$ to $\ket{g}$ relaxation is
$\gamma_\text{diss}= 1/\tau_\text{diss}$.  The excited state $\ket e$
relaxes to $\ket t$ with a rate of $\text{QY}_e \gamma_\text{diss}$
and to $\ket{g}$ with a rate of $(1-\text{QY}_e) \gamma_\text{diss}$.
Similarly, the relaxation rate from $\ket{t}$ to $\ket{p}$ is given by
$\text{QY}_t \gamma_\text{diss}$ and from $\ket{t}$ to $\ket{g}$ by
$(1-\text{QY}_t)\gamma_\text{diss}$.  These parameters are also
described in Table ~\ref{tab:controlparams}.

The light-matter coupling superoperator is given by
\begin{align}
  \field(t) \super V \rho = \field(t) \commute{\mu}{\rho} = E_0 \mu_0 E(t) \commute{u}{\rho} \label{eq:model_lm_1}
\end{align}
where $u=\mu/\mu_0$ and $E(t)=\field(t)/E_0$ are dimensionless.  The
parameter $E_0\mu_0$ tracks the perturbation strength and has units of
energy.  Matrix elements of $u$ are given in
Table~\ref{tab:controlparams}.  The field function $E(t)$ is given in
the frequency domain by
\begin{align}
  E(\omega) &= g(\omega; \omega_0, \sigma, \chi) + g(\omega; -\omega_0, \sigma, -\chi)
\end{align}
where the function $g(\omega; \omega_0, \sigma, \chi)$ is a
dimensionless, normalized Gaussian with  chirp $\chi$:
\begin{align}
  g(\omega; \omega_0, \sigma, \chi) &= \frac{1}{(2\sigma)^{1/2}\pi^{1/4} }
  \exp\left(-\frac{(\omega-\omega_0)^2}{2\sigma^2} - i \chi (\omega -\omega_0)^2\right)
\end{align}
The standard deviation of the field $\sigma$ in the Fourier domain is
computed from the FWHM in the time domain as,
\begin{align}
  \sigma = \frac{2\sqrt{2\log 2}}{\text{FWHM}_t} \label{eq:model_lm_4}
\end{align}
Parameters of the field are given in Table~\ref{tab:fieldparams}.

\begin{table}
\begin{ruledtabular}
  \begin{tabular}{llllllll}
          & $E_v$ & $E_t$ & $u_{gv}$ & $u_{ve}$ & $u_{et}$ & $u_{gt}$ & $u_{tp}$ \\
2 vs. 2   & 1.36  & 1.40 & 0.01      & 1.00      &  ---       &  ---      & ---        \\
1 vs. 3   & 0.8   & 1.35 & 0.01      & 1.00      &  1.00      &  0.0002   & ---        \\
Pump-dump & 0.8   & 1.40 & ---       & ---       &  ---       &  0.01   & 1.00       \\
\hline
          & $E_g$ & $E_e$     & $E_p$     & $u_{ge}$ &          &  & \\
All control scenarios       & 0.0   & 2.76      & 0.10      &  ---       &        & &    \\
          & $\tau_\text{el,deco}$ & $\tau_\text{vib,deco}$ & $\tau_\text{diss}$ & $\text{QY}_e$ & $\text{QY}_t$
          & \\
          & 90 fs & 500 fs            & 1000 fs            &  0.5          & 0.8           & &
\end{tabular}
\end{ruledtabular}
\caption{
  Parameters for each of the three control scenarios  corresponding
  to the three mechanisms described in the text.  Energies are in $eV$ while
  quantum yields $\text{QY}_t$ and $\text{QY}_e$ and transition
  strengths $u_{ij}$ are dimensionless.} \label{tab:controlparams}
\end{table}

\begin{table}
\begin{ruledtabular}
  \begin{tabular}{llll}
    & $\mu_0 E_0$ & $\hbar\omega_0$ & FWHM$_t$ \\
    Control scenarios$^\dagger$ & 0.05 eV & 1.38 eV & 80 fs\\
    Retinal/ChR2 model$^*$ &  0.047 eV & 1.25 eV & 120 fs
\end{tabular}
\end{ruledtabular}
{\footnotesize
  $\dagger$: $\mu_0 E_0$ is 0.20 eV for the pump-dump case.\\
  $*$ : Visible absorption (Fig.~\ref{fig:one_photon_prop}) computed using $\hbar\omega_0=2.5 $ eV.}
\caption{Parameters of the fields used throughout.} \label{tab:fieldparams}
\end{table}

\section{Two-photon control of retinal isomerization}\label{sec:model-retinal}
The two-photon photoisomerization of retinal is computed from the one
vibrational mode and two electronic state model of
Ref.~\onlinecite{balzer_mechanism_2003}.  The molecular Hamiltonian is
given by
\begin{align}
  H_0(\phi) = -\frac{1}{2I}\begin{pmatrix}
   \frac{\partial^2}{\partial \phi^2} & 0 \\
  0 & \frac{\partial^2}{\partial \phi^2}\end{pmatrix} + \begin{pmatrix}
    \alpha_1 (1-\cos \phi ) & \lambda \\
  \lambda & E_2 - \alpha_2 (1 - \cos\phi)\end{pmatrix}
\end{align}
where $\phi$ is a torsional mode with angles from 0 to $2\pi$.  The
\textit{trans} and \textit{cis} subspaces are defined by the following
projectors,
\begin{align}
  P_\text{trans}(\phi) &= \begin{cases}
    1 & \text{ for } |\phi|<\pi/2\\
    0 & \text{otherwise}
  \end{cases}\\
  P_\text{cis}(\phi) &= 1 - P_\text{trans}(\phi)
\end{align}
It should be noted that this is the opposite of the definitions of
Ref.~\onlinecite{balzer_mechanism_2003} since the ground state of
retinal in channelrhodopsin is the \textit{trans}
state.\cite{schneider_biophysics_2015}  To keep the number of
adjustable parameters to a minimum, only the \textit{cis} and
\textit{trans} labels from Ref.~\onlinecite{balzer_mechanism_2003}
were changed.  Two-photon absorption, c.f.
eqs.~(\ref{eq:mu})-(\ref{eq:mu_cross}), is obtained by adding a
near-resonant state at energy $E_v$.  Light-matter coupling is computed
with eqs.~\eqref{eq:model_lm_1}-\eqref{eq:model_lm_4} above.  The
transition operator in this case is given by
\begin{align}
  u = a_e \ket{g}\bra{e} +  a_v \ket{g}\bra{v} + \text{h.c.}
\end{align}
Eq.~\eqref{eq:mu_cross} is satisfied by setting $\mu_0 E_0 = 0.047$ eV
and the ratio $|a_v|^2/|a_e|^2$ to $1/86$.  Model parameters are given
Table~\ref{tab:retparams} and field parameters in
Table~\ref{tab:fieldparams}.

The retinal environment that stabilizes the \textit{cis} and
\textit{trans} isomers is evaluated in the secular Redfield
approximation\cite{weiss_quantum_2012} and integrated as part of the
Green's function calculation in the same manner as the Lindblad
tensors above.  The system part of the system-bath coupling is given by
the following dimensionless operator,
\begin{align}
  V_{sb}= \begin{pmatrix}\sqrt{1-y}(\cos \phi) & 0 \\ 0 & \sqrt{y}(1-\cos \phi) \end{pmatrix}
\end{align}
where the $\cos \phi$ part selects for the \textit{trans}
configuration, $1-\cos\phi$ for the \textit{cis} configuration and $y$
is a dimensionless parameter that determines the yield of the
photoisomerization.  The bath spectral density is ohmic with a cutoff,
given by
\begin{align}
  J(\omega) = \eta \omega e^{-\omega/\omega_c}.
\end{align}
The temperature of the Bose-Einstein bath is set to absolute zero.
This ensures that both the cis and trans states are steady states.
Indeed, this standard model does not otherwise reproduce the extremely
long-lived trans quasi-steady state of retinal in
channelrhodopsins,\cite{lorenz-fonfria_channelrhodopsin_2014} which
is critical to the repeated interaction amplification of control, as
described in eq.~(\ref{eq:timescales}).

\begin{table}
\begin{ruledtabular}
  \begin{tabular}{lllllll}
    Retinal & $I$   & $\alpha_1$& $\alpha_2$& $E_2$  & $\lambda$  & $E_v^\dagger$ \\
    & 4.84$\times 10^{-4}$ & 3.6 & 1.09 & 2.48 & 0.065 & 1.22\\
    Environment & $\eta$   & $\omega_c$& $y$$^{\dagger}$ & & & \\
                  & 0.45     & 0.035     & 0.8 & & &
\end{tabular}
\end{ruledtabular}
  {\footnotesize
    $\dagger$: Parameter modified from or not found in Ref.~\onlinecite{balzer_mechanism_2003}.}
  \caption{
    Parameters of the microscopic model of
    retinal\cite{balzer_mechanism_2003} used to compute two-photon
    activation rates.  The yield parameter $y$ is dimensionless; every
    other value is in eV.}
  \label{tab:retparams}
\end{table}

\bibliography{2photon_rhodopsin}

\end{document}